\documentclass[aps,reprint,amsmath,amssymb,showpacs,showkeys]{revtex4-1}
\usepackage{graphicx}
\usepackage{dcolumn}
\usepackage{bm}
\usepackage{hyperref}
\usepackage{natbib}  
\usepackage[justification=justified]{caption}
\usepackage{subcaption} 
\usepackage{amsopn}
\usepackage{float}
\usepackage{tikz}
\usetikzlibrary{shapes.geometric, arrows}
\tikzstyle{sqr2} = [rectangle, minimum width=3.5cm, minimum height=0.9cm, text centered, draw=black, fill=yellow!30]

\usepackage{booktabs} 

\newcommand{\Msun}{$\text{M}_{\odot}$}

\begin{document}
\graphicspath{{figures/}}	

\title[Stepping-stone sampling algorithm for evidence computation]{The stepping-stone sampling algorithm for calculating the evidence of gravitational wave models}

\author{Patricio Maturana Russel$^{1}$, Renate Meyer$^1$, John Veitch$^2$ and Nelson Christensen$^{3,4}$}
\affiliation{$^1$ Department of
  Statistics, University of Auckland, Auckland 1142, New Zealand \\
$^2$ Institute for Gravitational Research, School of Physics and Astronomy, University of Glasgow, Glasgow G12 8QQ, United Kingdom\\
$^3$ ARTEMIS, Universit\'{e} C\^{o}te d'Azur, Observatoire de la C\^{o}te d'Azur, CNRS, CS 34229, F-06304 Nice Cedex 4, France\\
$^4$ Physics and Astronomy, Carleton College, Northfield, MN 55057, USA}

\begin{abstract}
Bayesian statistical inference has become increasingly important for the analysis of observations from the Advanced LIGO and Advanced Virgo gravitational-wave detectors. To this end, iterative simulation techniques, in particular nested sampling and parallel tempering, have been implemented in the software library LALInference to sample from the posterior distribution of waveform parameters of compact binary coalescence events. Nested sampling was  mainly developed to calculate the marginal likelihood of a model but can produce posterior samples as a by-product. Thermodynamic integration is employed to calculate the evidence using samples generated by parallel tempering but has been found to be computationally demanding. Here we propose the stepping-stone sampling algorithm, originally proposed by Xie et al.\ (2011) in phylogenetics  and a special case of path sampling, as an alternative to thermodynamic integration.  The stepping-stone sampling algorithm is also based on samples from the power posteriors of parallel tempering but has  superior performance as fewer temperature steps and thus computational resources are needed to achieve the same accuracy.  We demonstrate its performance and computational costs in comparison to thermodynamic integration and nested sampling in a simulation study and a case study of computing the marginal likelihood of a binary black hole signal model applied to simulated data from the Advanced LIGO and Advanced Virgo gravitational wave detectors. To deal with the inadequate methods currently employed to estimate the standard errors of evidence estimates based on power posterior techniques, we propose a novel block bootstrap  approach and show its potential in our simulation study and LIGO application.
\end{abstract}

\pacs{04.30.-w, 02.50.-r, 05.45.Tp, 97.60.Bw}

\keywords{thermodynamic integration, stepping-stone sampling, nested sampling, model selection, gravitational waves}

\maketitle

\section{Introduction}

It has now been two decades since Bayesian parameter estimation routines were first introduced for studies in astrophysics~\cite{1994AJ....107.1295S}, gravitational waves~\cite{PhysRevD.58.082001}, and cosmology~\cite{0264-9381-18-14-306,1538-4357-563-2-L95}. Bayesian parameter estimation routines have become extremely important for these disciplines, and their use is ubiquitous~\cite{doi:10.1146/annurev-astro-082214-122339}. 
Recent dramatic observations with, for example, the cosmic microwave background and gravitational waves, have been used with Bayesian parameter estimation methods to significantly push our knowledge of the universe and its history~\cite{0067-0049-208-2-19, Planck2013,Planck2015, PhysRevLett.116.061102,PhysRevLett.116.241102,doi:10.1111/j.1740-9713.2016.00896.x,PhysRevLett.119.161101}. Advances in computer power, coupled with new and innovative Bayesian parameter estimation techniques, continue to push the applicability and importance of Bayesian methods~\cite{doi:10.1093/nsr/nwx044}.

The importance of accurate parameter estimation calculations was dramatically displayed with the observations of gravitational waves and gamma rays from the binary neutron star merger GW170817 and GRB 170817A~\cite{PhysRevLett.119.161101,2041-8205-848-2-L14}. Using the data from the two Advanced LIGO detectors~\cite{0264-9381-32-7-074001} and the Advanced Virgo detector~\cite{0264-9381-32-2-024001} an initial sky-map and distance estimates from the gravitational wave data was released five hours after the merger~\cite{2041-8205-848-2-L12} using a specially designed method for sky position estimation~\cite{PhysRevD.93.024013}. A little over 11 hours after the gravitational wave - gamma ray event refined estimates were released based on the first comprehensive parameter estimation~\cite{Veitch:2015}, giving a more accurate estimated of the sky position and distance to the source~\cite{2041-8205-848-2-L12}. The parameter estimation calculations allowed astronomers to identify the source, providing for electromagnetic observations that yielded a plethora of astrophysical information, including the observation of a kilonova~\cite{2041-8205-848-2-L12}. Parameter estimation of gravitational wave models  will continue to be significant  for multimessenger astronomy.

Also of critical importance is the ability to conduct model comparison and parameter estimation studies with the gravitational wave signals. For example, Bayesian parameter estimation methods were used to decipher the physical characteristics of the observed gravitational wave events, such as the first observed binary black hole merger GW150914~\cite{PhysRevLett.116.061102,PhysRevLett.116.241102,PhysRevX.6.041014,O1BBH}, and the binary neutron star merger GW170817~\cite{PhysRevLett.119.161101,Abbott:2018wiz}. Similarly, model comparisons have been conducted in a number of ways using the data from the detected gravitational waves signals. This includes tests of general relativity~\cite{PhysRevLett.116.221101}, neutron star equation of state studies~\cite{PhysRevLett.119.161101,Abbott:2018exr}, constraining tidal instabilities in binary neutron star mergers~\cite{Weinberg:2018icl}, and the search for a stochastic gravitational wave background from binary black hole mergers over the history of the universe~\cite{PhysRevX.8.021019}. When Advanced LIGO and Advanced Virgo made the first observation of a binary black merger using the data from three detectors it provided an opportunity to conduct a model comparison test as to whether the polarization of the gravitational waves was consistent with general relativity, or other theories of gravity; general relativity succeeded in this important model comparison~\cite{PhysRevLett.119.141101}. Advanced LIGO data was also used to search for a stochastic gravitational wave background as described by general relativity or alternative theories of gravity, and model comparison was integral to this study~\cite{PhysRevX.7.041058,PhysRevLett.120.201102}. Methods that  improve the calculation of the evidence, the marginal likelihood of a model, would be well-received in the gravitational wave community, and would certainly be of use in other areas of astrophysics and cosmology~\cite{doi:10.1146/annurev-astro-082214-122339}. Equally important is an accurate estimation of the associated standard error. Here we introduce the moving block bootstrap (MBB) that accounts for the autocorrelation between the samples, and provides a more accurate estimate than the standard bootstrap method.


Presented in this paper is the stepping-stone sampling (SS) algorithm that provides an improvement of the evidence estimator for Bayesian model selection and the MBB for computing its standard error. The practicality of the SS algorithm for calculating the evidence of gravitational wave models will be demonstrated. The SS algorithm could be a further advancement for model selection for gravitational waves data analysis, as well as for other applications in astrophysics and astronomy.













The paper is structured as follows. In Section II we review nested sampling, thermodynamic integration and introduce the stepping-stone algorithm for computing the evidence of Bayesian model selection. In Section III we introduce the moving block bootstrap for calculating the Monte Carlo standard error of the evidence estimates. In Section IV we show the enhanced performance of the SS algorithm over thermodynamic integration in a simulation study. The different algorithms are then  applied  to simulated LIGO-Virgo gravitational wave data in Section V. Their results are contrasted and  the benefits of the SS  algorithm and the moving black bootstrap for standard error estimation become evident. A summary discussion is given in Section VI.

\section{Computation of Marginal Likelihood}

The \textit{evidence} or \textit{marginal likelihood} of a model  $M$ is a multi-dimensional integral defined as
\begin{align}
\label{eq:z}
z = \int_{\Theta}L(\bm{X}|\bm{\theta},M)\pi(\bm{\theta}|M) \text{d} \bm{\theta},
\end{align}
where $\bm{\theta} \in \Theta$ denotes the parameter vector, $\bm{X}$ the dataset, $L(\bm{X}|\bm{\theta},M)$  the likelihood function, and $\pi(\bm{\theta}|M)$  the prior density, assumed to be proper, i.e.\ $ \int_{\Theta}\pi(\bm{\theta}|M) \text{d} \bm{\theta}=1$.

In general, this integral (\ref{eq:z}) has no analytical solution and must be estimated using numerical methods.  Importance sampling techniques, in particular the arithmetic mean (AM) and harmonic mean (HM) methods,  provide the simplest way of estimating it \citep{Newton:Raftery:1994}.
Let $\bm{\theta}_i, i=1,\ldots,n$ be samples from the prior, the AM estimator is an average of corresponding $n$ likelihood values:
\begin{align}
\label{eq:AM}
\widehat{z}_{AM} = \frac{1}{n}\sum_{i=1}^n L(\bm{X}|\bm{\theta}_i,M).
\end{align}
In general, high-likelihood areas are very small and  constitute a small fraction of the prior.  Therefore, unless $n$ is very large, the sample will not adequately represent these areas and yield a poor estimate.  The HM estimator is  based on samples 
$\bm{\theta}_i, i=1,\ldots,n$
 drawn from the posterior:
\begin{align}
\label{eq:HM}
\widehat{z}_{HM} = \left( \frac{1}{n}\sum_{i=1}^n \frac{1}{ L(\bm{X}|\bm{\theta}_i,M)} \right)^{-1} ~ .
\end{align}
This is the harmonic mean of likelihood values.

The AM and HM estimators are not recommended because they produce unreliable estimates of the evidence, even though they are easily calculated.   In this context, more complex approaches have been proposed, such as power posterior methods \cite{Xie:Lewis:Fan:Kuo:Chen:2011, Lartillot:Philippe:2006, Friel:2008, Neal:2001}.  These methods rely on a set of transitional distributions which connect the prior and the posterior, reminiscent of simulated annealing.  The geometric path is the most popular scheme used to connect these distributions and defines the \textit{power posterior density} as
\begin{align}\label{eq:powerposterior}
p_{\beta}(\bm{\theta}|\bm{X}, M) = \frac{L(\bm{X}|\bm{\theta},M)^{\beta}\pi(\bm{\theta}|M)}{z_{\beta}},
\end{align}
for the inverse temperature $0\leq\beta\leq 1$, where $z_{\beta}$ is the normalizing constant, which is defined as $\int_{\Theta} L(\bm{X}|\bm{\theta},M)^{\beta}\pi(\bm{\theta}|M) \text{d}\bm{\theta}$. Note that the power posterior density turns into the prior and posterior for $\beta = 0$ and $\beta = 1$, respectively.   

Methods  that make use of samples from the power posteriors are much more accurate than HM as has been widely documented \citep{Lartillot:Philippe:2006, Newton:Raftery:1994, Xie:Lewis:Fan:Kuo:Chen:2011}, particularly in high dimensional problems.  Among these methods, thermodynamic integration (TI) \citep{Lartillot:Philippe:2006} is a popular method to estimate the  evidence  of gravitational wave (GW) models, showing in general good performance.  Another method, widely applied in other fields such as phylogenetics is the SS algorithm \citep{Xie:Lewis:Fan:Kuo:Chen:2011}.  As this method
can provide many advantages over the TI estimate, it is important to explore the performance of  the SS estimator  for GW models as  to the best of our knowledge,  the SS algorithm has not been used for evidence calculation in this context.

One of the drawbacks of power posterior methods is the significant computational cost required to produce a single evidence estimate as 
multiple Markov chains have to be run, one for each temperature.  Fortunately, since  parallel tempering is commonly used in GW parameter estimation, the samples at different temperatures are available and can be recycled in order to use these methods.

However, as has been noticed in \cite{Veitch:2015}, TI might require a larger number of temperatures than the one needed for parameter estimation in order to achieve accurate estimates.  Note that the samples of chains  at  temperatures $T>1$ ($\beta<1$) are only used to aid the mixing of  the chain at $T=\beta=1$ whose stationary distribution is the posterior, and  are therefore discarded from the inference process.  In this context, the SS algorithm seems very promising since  it requires fewer temperature steps than TI to provide accurate evidence estimates as we will show in section \ref{sec:simulation}.

Another method to estimate the evidence,  not based on power posteriors, is nested sampling (NS) \citep{Skilling:2006, Veitch:2010}.  This Bayesian algorithm has been successfully applied in diverse fields, such as astronomy \citep{Brewer:Donovan:2015}, cosmology \citep{Feroz:2009}, engineering \citep{Henderson:2017} and phylogenetics \citep{Maturana:2017b, Maturana:2018}. To estimate the evidence of GW models, NS has been implemented in the software package LALInference \citep{Veitch:2015}.  The method has the unique property of yielding an estimation of the uncertainty associated with the evidence estimate in a single run (however, only for independent samples).

Alternatively, instead of estimating the evidence for each model being tested, a trans-dimensional Reversible Jump Markov chain Monte Carlo \citep[RJMCMC;][]{Green:1995, Umstatter:2005} method can be used in order to explore the joint space of all models.  Then the probability for each model can be calculated simply by calculating the relative frequency of visits to each model by the Markov chain.  However, this exploration depends on tuning parameters which can be difficult to specify, leading to poor mixing of the Markov chain and subsequently to large statistical errors associated with the evidence estimates \citep{Cornish:2014}.


Below we describe TI, SS and NS in more detail before comparing their performance in sections \ref{sec:simulation} and \ref{sec:LIGO}.

\subsection{Thermodynamic Integration}
Thermodynamic integration or the more general path sampling \citep{Gelman:1998} make use of an auxiliary variable $\beta$, $0\leq \beta \leq 1$, to
define transitional distributions, namely the power posterior distributions defined in (\ref{eq:powerposterior}) in the case of TI, that provide a path from the prior ($\beta=0$) to the posterior distribution ($\beta=1$). By explicitly denoting the evidence $z_\beta$ as a function of $\beta$ by
\begin{align}
z(\bm{X}|\beta)=\int_{\Theta} L(\bm{X}|\bm{\theta},M)^{\beta}\pi(\bm{\theta}|M) \text{d}\bm{\theta},
\end{align}
the log marginal likelihood has the representation as the integral over the 1-dimensional parameter $\beta$ of half the mean deviance where the expectation is taken with respect to the power posterior:
\begin{equation}\label{eq:TI}
\log(z)=\log\left( \frac{z(\bm{X}|\beta=1)}{z(\bm{X}|\beta=0)}\right)=\int_0^1 E_{\beta} \left[ \log(p(\bm{X}|\bm{\theta},M)\right]\text{d}\beta.
\end{equation}
Representation (\ref{eq:TI}) follows by integration from
\begin{align*}
&\frac{\partial}{\partial\beta} \log(z(\bm{X}|\beta)) = \frac{1}{z(\bm{X}|\beta)} \frac{\partial}{\partial\beta} z(\bm{X}|\beta)\\
&= \frac{1}{z(\bm{X}|\beta)} \frac{\partial}{\partial\beta}\int_{\Theta} L(\bm{X}|\bm{\theta},M)^{\beta}\pi(\bm{\theta}|M) \text{d}\bm{\theta}\\
&=  \frac{1}{z(\bm{X}|\beta)} \int_{\Theta}  L(\bm{X}|\bm{\theta},M)^{\beta}\log(L(\bm{X}|\bm{\theta},M)) \pi(\bm{\theta}|M) \text{d}\bm{\theta}\\
&=  \int_{\Theta}   \frac{L(\bm{X}|\bm{\theta},M)^{\beta}\pi(\bm{\theta}|M)}{z_{\beta}} \log(L(\bm{X}|\bm{\theta},M))\text{d}\bm{\theta}\\
&= E_{\beta} \left[ \log(L(\bm{X}|\bm{\theta},M)\right].
\end{align*}
The samples from the parallel tempered chains for different values of $\beta$ provide samples from the power posteriors and the
expectation $E_{\beta}\left[ \log(L(\bm{X}|\bm{\theta},M)\right]$ is then estimated by the sample average. The integral
in equation (\ref{eq:TI}) is then approximated by numerical integration, e.g. using the trapezoidal or Simpson's rule.

\subsection{Stepping-stone Sampling Algorithm}
Stepping-stone sampling is another method to estimate the marginal likelihood.  It has been widely used by the phylogenetic community where it was proposed by \cite{Xie:Lewis:Fan:Kuo:Chen:2011}.  SS works basically by mixing elements from importance sampling and simulated annealing methods.  This method relies on the same sampling scheme required by TI.  Therefore, its implementation in any software package where TI or parallel tempering has already been implemented should be straightforward.  SS has the advantage of requiring fewer path steps than TI to accurately estimate  the marginal likelihood and yielding a less-biased estimator as demonstrated in section \ref{sec:simulation}.

The marginal likelihood can be seen as the ratio $z = z_1/z_0$, where $z_0 = 1$ since the prior is assumed to be proper.  The direct calculation of this ratio via importance sampling is not reliable because the distributions involved in the numerator and denominator (posterior and prior, respectively) are, in general, quite different.  To solve this problem, SS expands this ratio in a telescope product of $K$ ratios of normalizing constants of the transitional distributions \cite{Neal:1993}, that~is
\begin{align*}
\label{eq:SS_ratios}
z = \frac{z_{1}}{z_{0}} = \frac{z_{\beta_{1}}}{z_{\beta_{0}}}\frac{z_{\beta_{2}}}{z_{\beta_{1}}} \dots
\frac{z_{\beta_{K-2}}}{z_{\beta_{K-3}}} \frac{z_{\beta_{K-1}}}{z_{\beta_{K-2}}} = \prod_{k=1}^{K-1} \frac{z_{\beta_{k}}}{z_{\beta_{k-1}}} = \prod_{k=1}^{K-1} r_k,
\end{align*}
for $\beta_0 = 0 < \beta_1 < \dots < \beta_{K-2}<\beta_{K-1} =~1$, being the sequence of inverse temperatures, where $r_k=z_{\beta_k}/z_{\beta_{k-1}}$.  These individual intermittent ratios can be estimated  with higher accuracy than
$\frac{z_1}{z_0}$ because the  distributions in the
numerator and denominator are generally quite similar when using a reasonable number of temperatures $K$.  In this situation the importance sampling method works well. 

SS estimates each ratio $r_k$ by importance sampling using $p_{\beta_{k-1}}$ as importance sampling distribution.  This is a suitable distribution because it has heavier tails than $p_{\beta_{k}}$ which leads to an efficient estimate of $r_k$.  In this manner, it avoids estimating from the posterior distribution, making it slightly less expensive computationally than TI for the same number of path steps.  The estimation of each ratio is based on the identity
\begin{align*}
r_k = \frac{z_{\beta_{k}}}{z_{\beta_{k-1}}} &= \int_{\Theta}\frac{L(\bm{X}|\bm{\theta},M)^{\beta_{k \quad}}}{L(\bm{X}|\bm{\theta},M)^{\beta_{k-1}}} \: p_{\beta_{k-1}}(\bm{\theta}|\bm{X},M) \text{d}\bm{\theta}, 
\end{align*}
which is estimated by its unbiased Monte Carlo estimator
\begin{align*}
\widehat{r}_k = \frac{1}{n} \sum_{i = 1}^{n}  L( \bm X | \bm{\theta}_{\!\beta_{k-1}}^{i}, M)^{\beta_{k}-\beta_{k-1}},
\end{align*}
where $\bm{\theta}_{\!\beta_{k-1}}^{1}, \dots, \bm{\theta}_{\!\beta_{k-1}}^{n}$ are drawn from $p_{\beta_{k-1}}$ with $k = 1, \dots, K-1$. 

Therefore,  the SS estimate of the marginal likelihood is defined as 
\begin{align*}
\widehat{z} &= \prod_{k=1}^{K-1}\frac{1}{n} \sum_{i = 1}^{n}  L( \bm X | \bm{\theta}_{\!\beta_{k-1}}^{i}, M)^{\beta_{k}-\beta_{k-1}},
\end{align*}
with log-version
\begin{align*}
\log\widehat{z} &= \sum_{k=1}^{K-1} \log \sum_{i = 1}^{n}  L( \bm X | \bm{\theta}_{\!\beta_{k-1}}^{i}, M)^{\beta_{k}-\beta_{k-1}} - (K-1) \log n.
\end{align*}
Although $\widehat{z}$ is unbiased, the log transformation introduces a bias which can be alleviated by increasing $K$ \citep{Xie:Lewis:Fan:Kuo:Chen:2011}.

The performance of this method depends naturally on its specifications such as the number of transitional distributions and number of samples from each of them ($K$ and $n$, respectively).  The dispersal of the $\beta$ values has also a strong influence, even more so in TI (see \cite{Xie:Lewis:Fan:Kuo:Chen:2011} and our simulation study below).  Along these lines, \cite{Xie:Lewis:Fan:Kuo:Chen:2011} proposed to spread the $\beta$ values according to the evenly spaced quantiles of a Beta(0.3, 1) distribution.  This distribution is right skewed, thereby putting half of the $\beta$ values below 0.1 where most of the variability  is found.    

SS is closely related to annealed importance sampling \citep[][]{Neal:2001}.  The latter utilizes the same product of ratios, but instead of estimating each ratio separately, it estimates the entire product via importance sampling, that is the whole telescope product is evaluated multiple times and then these values are averaged \citep{Maturana:2017}.  For the particular case of $K = 2$, that is considering only the prior, both methods reduce to the arithmetic mean, and for $n =1$, they are equivalent.

\subsection{Nested Sampling}

NS transforms the multidimensional integral defined in~\eqref{eq:z}, by making use of a property of positive random variables (see~\cite{Maturana:2017b} for more details), into a one-dimensional one that utilizes a function that relates the prior with the likelihood as 
\begin{align*}
z = \int_{0}^{1} L(\xi)\text{d}\xi, 
\end{align*}
where $L$ is the likelihood as a function of the prior volume~$\xi$.  This function can be read as the proportion of prior volume $\xi$ with likelihood values greater than $L(\xi)$. 

This likelihood is a non-increasing function over the unit range.  For a given decreasing sequence of $\xi$-values and an increasing sequence of $L$-values, the marginal likelihood can be estimated using, for instance, the trapezium rule
\begin{align*}
\widehat{z}_{NS} = \sum_{i=1}^{K} \dfrac{1}{2}(\xi_{i-1} - \xi_{i+1})L_i,
\end{align*}
where $0 < \xi_{K+1} < \xi_{K} < \cdots < \xi_1<\xi_0 = 1$.

NS explores the parameter space from the prior toward those areas of high likelihood values over time.  For this, a set of $N$ points, called \textit{live} points, are drawn independently from the prior.  The point $\bm{\theta}_1$ with the lowest likelihood associated to these points is detected and the latter is registered as $L_1$.  Then, this point $\bm{\theta}_1$ is replaced by a new one $\bm{\theta}^*$ drawn from the prior but restricted to have a greater likelihood, that is $L(\bm{\theta}^*) > L(\bm{\theta}_1)$.  This procedure is repeated until a given stopping criterion is satisfied.  Thus, an increasing sequence of likelihood values $L_1, \dots, L_K$ is generated.

Even though the $\xi$-values cannot be measured precisely, the nature of this algorithm allows them to be estimated. 
The $\xi$-sequence can be defined as 
\begin{align*}
\xi_1 = u_1,\: \xi_2 = u_2 \xi_1, \dots,\: \xi_K = u_K \xi_{K-1},
\end{align*}
where $u_i\sim\text{Beta}(N,1)$.  The geometric mean is the most common method to estimate the $u$-values, which yields
\begin{align*}
\xi_i = e^{-i/N}.
\end{align*}

The nature of NS algorithm also allows to estimate the standard error of the $\log z$ estimate in a single run as 
\begin{equation}\label{eq:SDNS}
\widehat{s.e.}_{\text{NS}}(\log z) = \sqrt{\dfrac{H}{N}},
\end{equation}
where $H$ is the negative entropy.  However, this NS standard error estimate is only valid if the samples are drawn independently. In practice though, the samples will often be serially dependent because Metropolis-Hastings algorithms are  used for their generation. As an alternative to (\ref{eq:SDNS}), for a fixed sequence of likelihood values and multiple sequence of $\xi$-values, generated from different  $u\!\sim\!\text{Beta}(N,1)$ values, a distribution of marginal likelihood estimates can be generated and subsequently the uncertainty can be estimated. 
\bigskip

\section{Estimation of the Monte Carlo Standard Error of the Evidence}
The point estimate of the evidence is subject to random errors and therefore we need 
 to have a measure of the Monte Carlo standard error of the evidence estimates.  This is also important if we want to compare the  performance of different types of evidence estimates.  In the NS case, the algorithm provides direct ways of calculating its standard error from  a single run as given in (\ref{eq:SDNS}).  However, power posterior methods lack a reliable direct way of calculating the standard error of the evidence.  In \cite{Lartillot:Philippe:2006} and \cite{Xie:Lewis:Fan:Kuo:Chen:2011}, the authors proposed estimates which rely on the  independence of the samples in the Markov chains at different temperatures, an assumption that is not met  in general.  Practitioners opt for the standard procedure of repeating the analysis multiple times and then calculating the standard error.  This brute force technique can be very costly and is in some cases computationally not viable.  Alternatively, some estimate the error internally in a single run, that is by re-sampling independently the Markov chains in order to generate multiple evidence estimates.  However, this approach does not consider the potential autocorrelation in the samples, leading to wrong estimates.  Here, we propose the use of a block bootstrap method for multivariate time series, which accounts for  the autocorrelation between the samples within a Markov chain at a fixed
temperature and the cross-correlation between parallel chains at different temperatures.
	

Bootstrap is a resampling procedure proposed by \cite{Efron:1979}, initially for independent variables and later generalized by several authors.  An extension for the case of time series was proposed in \cite{Kunsch:1989}, which differs from the original algorithm by allowing the sampling in blocks.  The method is known as \textit{moving block bootstrap}, in short MBB.  This allows to take into account the presence of dependence in the data.

Let $X_1, \dots, X_n$ be the observed values from a sequence of stationary random variables, in our case, a Markov chain.  Define the overlapping blocks  $B_i = (X_i, ,\dots,X_{i+\ell-1})$ of length $\ell$, for $1\leq i \leq n-\ell + 1$ and $1 \leq \ell \leq n$, that is
\begin{align*}
B_1 &= (X_1, X_2, X_3, \dots, X_{\ell}) \\
B_2 &= (X_2, X_3, X_4, \dots, X_{\ell + 1}) \\
&\:\:\vdots\\
B_m &= (X_{n-\ell +1}, \dots, X_{n}), 
\end{align*} 
where $m = n - \ell +1$.  MBB works by resampling randomly $b$ blocks (for didactic reasons, suppose that $b = n/\ell$) and concatenating them in order to form a set of bootstrap observations $X^*_1,\dots,X^*_n$.  For $\ell = 1$, the original bootstrap method for i.i.d. data is recovered.  This procedure is repeated as usual, generating the distribution of the statistic of interest, in our case the marginal likelihood.  In the general case that $n$ is not a multiple of $\ell$, we can concatenate the random sample of $b$ block bootstraps, where $b$ is $n/\ell$ rounded up, and discard the leftover points $X^*_{n+1}, \dots, X^*_{b\ell}$, such that the bootstrap observation set has length $n$, as the original dataset.

Variants of this method can be found in \cite{Lahiri:2003}, such as \textit{stationary bootstrap}, where the block length follows a geometric distribution;  \textit{nonoverlapping block bootstrap}, which as its name say, considers nonoverlapping blocks; and \textit{circular block bootstrap}, which increases the original dataset with the first $\ell - 1$ observations in order to give equal weights to all of them. 

In the context of parallel tempering, in which case there are multiple Markov chains, we need to generate the bootstrap observations using the same scheme for all the chains.  For instance, assuming equal chain lengths, a bootstrap observation set for a Markov chain consisting in $(B_6,B_4,B_2)$, is replicated across the other chains.  This procedure takes into account the potential autocorrelation within the chains and the cross-correlation between the chains due to the swaps in parallel tempering sampling.  This is the approach applied in our examples.

\section{Simulation Study} \label{sec:simulation}
We consider a simple Gaussian model used by \cite{Lartillot:Philippe:2006} to test TI and compare it to the harmonic mean method.  Here, it is used to compare SS to TI.  We also assess the error estimate via the MBB method and compare it to the empirical calculation of the error.  In addition, we study NS performance for different sampling specifications.  

The model is parametrized by a vector $\bm{x} = (x_1,x_2, \dots, x_d)$ of dimension $d$. The prior on $\bm x$ is a product of independent standard normal distributions on each $x_i$, for $i = 1, \dots,d$.  The likelihood is 
\begin{align*}
L(\bm{x}) = \prod_{i=1}^{d} e^{- \frac{x_i^2}{2v}},
\end{align*}
where $v$ is a parameter.  Doing some calculations, it is easy to see that the posterior distribution is given by a product of independent $\text{N}(0, v/(1+v))$ distributions, and therefore, its marginal likelihood has an analytical solution, which is $z = (v/(1+v))^{d/2}$.  The power posterior or transitional distributions are given by a product of independent  $\text{N}(0,v/(v+\beta))$ distributions.  All the involved distributions are Gaussians, so the sampling required to calculate TI and SS is straightforward.  However, we use the Metropolis algorithm to sample these densities and thus allow a certain degree of autocorrelation in the samples, making the analysis more realistic in an evidence estimation context.  The Markov chains have a lag of around 18 on average.  In addition, we consider independent samples to assess MBB performance in the context of error estimation.  

\begin{figure}[]
	\begin{subfigure}{0.50\textwidth}
		\centering 
		\includegraphics[scale=0.45,clip=true,angle=0]{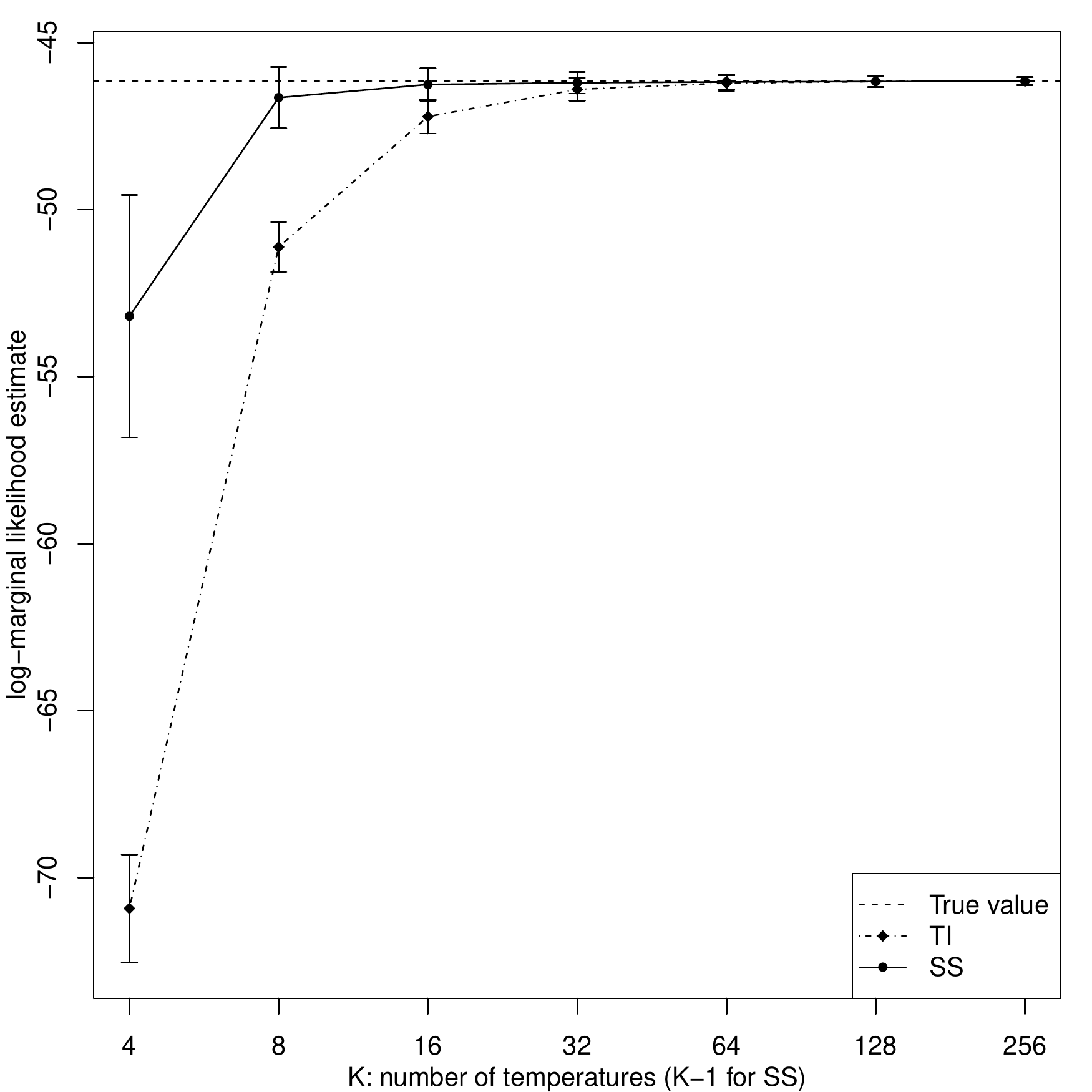}
		\caption{$\beta$ values spread according to evenly spaced quantiles of a Beta(0.3, 1) distribution.}
		\label{fig1a}
	\end{subfigure}
	\begin{subfigure}{0.50\textwidth}
		\centering 
		\includegraphics[scale=0.45,clip=true,angle=0]{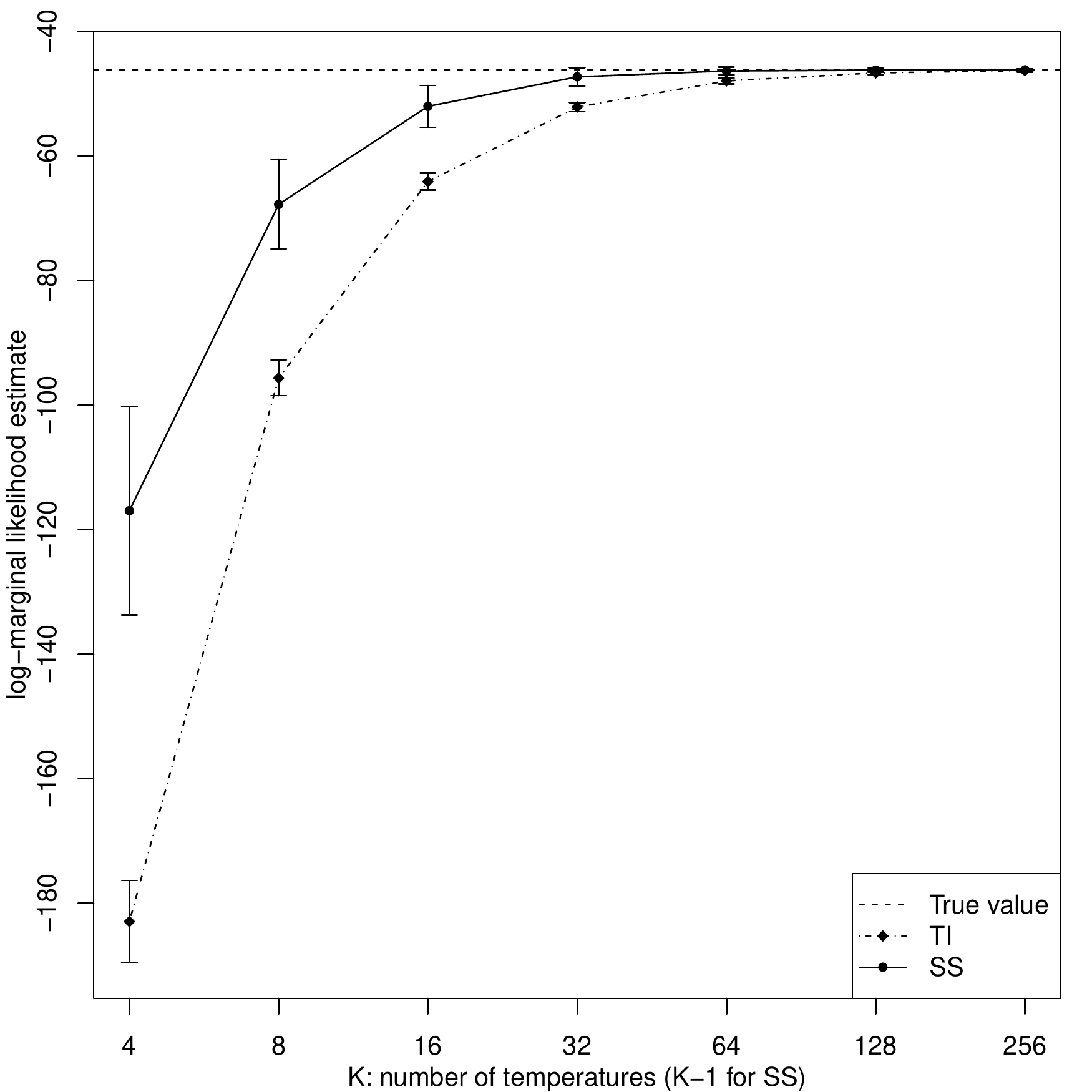}
		\caption{$\beta$ values equally spaced between 0 and 1.} 
		\label{fig1b}
	\end{subfigure}	
	\caption{Log-marginal likelihood estimates as a function of the number of temperatures $K$ for the Gaussian model.  Error bars depict $\pm1$ standard error based on 1000 independent MCMC analyses.}		
	\label{fig1}
\end{figure}

\subsection{Evidence estimate}

We consider the following model specifications: $v = 0.01$ and $d = 20$.  This yields a log-marginal likelihood value $-46.15$.  The analysis is performed for $n = 1000$ and $K = 4, 8, 16, 32, 64, 128, 256$.  Strictly speaking, SS uses $K-1$ temperatures, since it does not require samples from the posterior.  For the arrangement of the $\beta$ values, we test two approaches: evenly spaced values from 0 to 1, and values spread according to evenly spaced quantiles of a $\text{Beta}(0.3, 1)$ distribution.  The MCMC analysis is replicated 1000 times (with different random seeds) in order to calculate the error associated with the estimates.  The same power posterior samples are used to estimate SS and TI. 

Figures~\ref{fig1a} and \ref{fig1b} display the results. It becomes clear in both cases that the SS algorithm requires less temperatures than TI to produce estimates around the true value.  When the $\beta$ values are calculated according to a Uniform(0,1), Figure~\ref{fig1b}, the TI estimates are seriously biased for low number of temperatures, whereas the SS estimates, even though biased too,  are closer to the true value.  For equally spaced $\beta$ values and $K=4$ Figure~\ref{fig1b}, TI is more than 130 units away from the true value compared to the around 25 units for $\beta$ values spread according to quantiles of the Beta(0.3,1) distribution  in Figure~\ref{fig1a}.  This shows that TI is more sensitive to the distribution of the temperatures as was similarly shown by \cite{Xie:Lewis:Fan:Kuo:Chen:2011}.

Both methods improve their performance when most of the computational effort is allocated in sampling in power posterior distributions near the prior, that is for high temperatures.  This is the effect of the Beta(0.3, 1) distribution, which allows that half of the $\beta$ values are less than 0.1.  The results for this case are displayed in Figure~\ref{fig1a}.  Even though TI improves its performance considerably, it can not outperform SS, which still needs fewer step temperatures to produce estimates around the true value. 

\subsection{Standard error estimate}
\label{subsec:SDest}

Based on the case that the $\beta$ values follow a Beta(0.3, 1) distribution, we study the performance of the MBB method for estimating the evidence error.  For this, we calculate the standard error from the 1000 independent evidence estimates used in the previous analysis, call this $\widehat{\mbox{s.e.}}_{\text{ind}}$ and compare it to the standard error estimates calculated using MBB ($\widehat{\mbox{s.e.}}_{\text{MBB}}$ for different block lengths, $\ell = 1, 10, 30, 50, 100, 200, 300$, via their differences, $\widehat{\mbox{s.e.}}_{\text{MBB}}-\widehat{\mbox{s.e.}}_{\text{ind}}$.  

\begin{figure}[]
	\begin{subfigure}{0.50\textwidth}
	\centering 
	\includegraphics[scale=0.45,clip=true,angle=0]{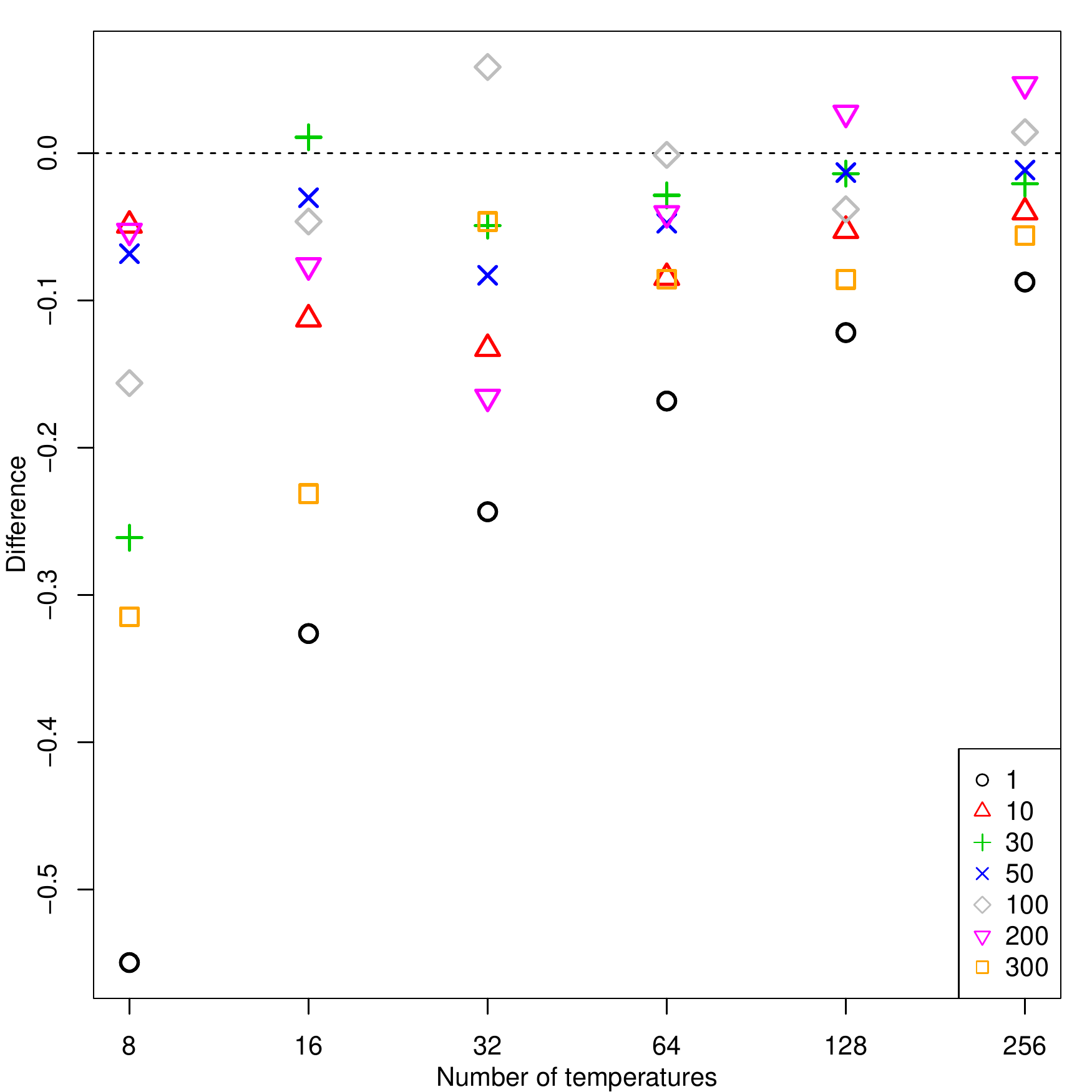}
	\caption{The Markov chains contain a degree of autocorrelation.}
	\label{fig2a}
	\end{subfigure}
	\begin{subfigure}{0.50\textwidth}
	\centering 
	\includegraphics[scale=0.45,clip=true,angle=0]{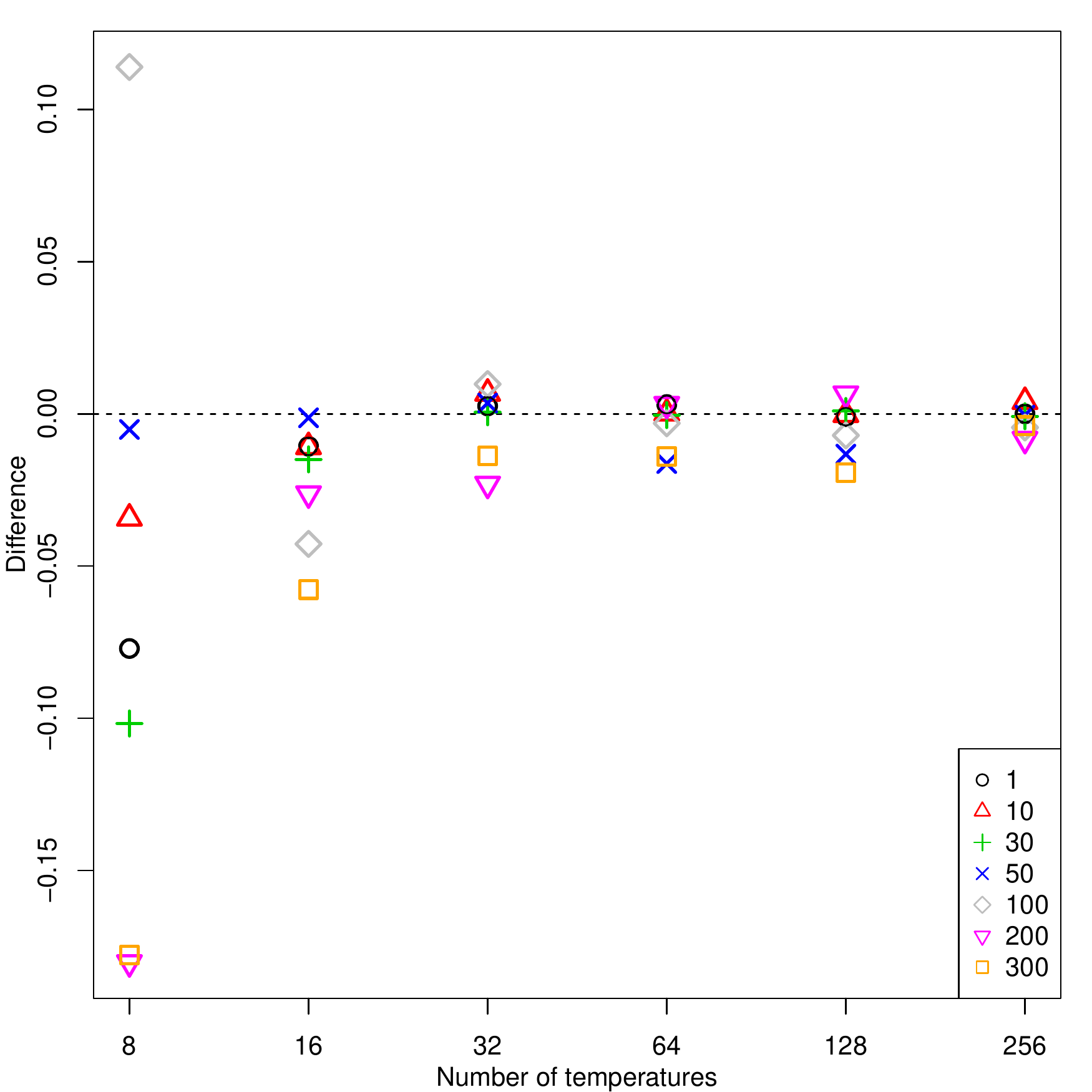}
	\caption{The samples in the Markov chain are completely independent.}
	\label{fig2b}
	\end{subfigure}
	\caption{Difference between the standard error calculated via MBB and the one from independent evidence estimates.  The legend shows the different block lengths used in MBB.}
	\label{fig2}
\end{figure}

The results are shown in Figure~\ref{fig2a}.  The case $\ell = 1$ is the original bootstrap method, which is used frequently for power posterior methods, but which ignores the dependence in the sampled values of the Markov chain.  It is obvious that in the simple bootstrap with block length $\ell=1$, the standard error is severely underestimated. On the other hand, the standard error estimates  improved significantly using 
the MBB with larger block lengths, but still some underestimate  the standard error.  However, this example is an extreme case of highly correlated Markov chains.  

We have also performed the analysis in the ideal case that the samples in the Markov chains are completely independent.  The result are displayed in Figure~\ref{fig2b}.  In this case, the standard bootstrap method, that is $\ell = 1$, is sufficient to estimate the standard error reasonably well. Large block lengths cause, in general, a slight underestimation but only in the case of a low number of temperatures.  As the number of temperatures increases, the estimates are located around the empirical error estimates, i.e., around zero, and less dispersed.


We caution against the use of the theoretical standard error estimate of NS in Equation (\ref{eq:SDNS}) when the Metropolis-Hastings algorithm is used to generate the samples rather
than sampling independently, as the validity of this theoretical standard error estimate is based on the independence assumption. To this end, we include a comparison of
this theoretical NS standard error estimate with the empirical standard error obtained from 100 independent runs in Table I.
We observe a decrease in bias with increasing number of MCMC steps. However,  the NS standard error estimates are far too small and thus underestimate the uncertainty
even for a large number of MCMC steps of 5000. This is a well known shortcoming of the NS standard error estimate, e.g.\ a more detailed examination of this issue can also be found in Figure 4 of  \cite{Veitch:2010}.

\begin{table}
	\caption{\label{tab:NSresults} Nested sampling results based on 100 independent NS runs.  $N$ is the number of live points, ``Steps" the number of MCMC steps used to generate the points at each iteration, $Ave(\widehat{s.e.}_{\text{NS}})$ the average of the theoretical standard error estimate defined in \eqref{eq:SDNS}, SD($\widehat{s.e.}_{\text{NS}}$) the standard deviation of the theoretical standard error estimates, $\widehat{s.e.}_{\text{NS,ind}}$ the standard error estimate based on the independent marginal likelihood estimates, and ``Bias" the difference between the true value and the mean of the NS marginal likelihood estimates.}
	\begin{ruledtabular}
		\begin{tabular}{cccccc}
			$N$ & Steps & $Ave(\widehat{s.e.}_{\text{NS}})$ & SD($\widehat{s.e.}_{\text{NS}}$) & $\widehat{s.e.}_{\text{NS,ind}}$  & Bias \\ \hline
			10&10&1.9922&0.0921&3.6244&4.7724\\
			10&100&1.8918&0.0562&2.3078&-0.3310\\
			10&1000&1.8959&0.0542&2.1083&-0.1580\\
			10&5000&1.8939&0.0562&2.3454&-0.3233\\
		\end{tabular}
	\end{ruledtabular}
\end{table}

\section{Application with Simulated LIGO-Virgo Data} \label{sec:LIGO}
We apply the SS algorithm to an example analysis of a simulated binary black hole
coalescence signal in the Advanced LIGO~\cite{0264-9381-32-7-074001} and Advanced
Virgo~\cite{0264-9381-32-2-024001}
gravitational wave detectors, operating at design sensitivity.
The data contained 4\,s of simulated Gaussian noise, generated using the design
sensitivity curves of two Advanced LIGO detectors (Hanford, Livingston) and the Advanced Virgo detector,
plus the GW signal. The simulated black hole binary had component masses
25\,\Msun and 13\,\Msun, and lay at a luminosity distance of 614\,Mpc, with a
combined signal-to-noise ratio of 17.9 in the three-detector network.
The analysis was performed in the frequency range 40--512\,Hz using the IMRPhenomPv2
waveform approximant~\cite{Hannam:2013oca}. The system's total angular momentum
was inclined at 95$^\circ$ to the line-of-sight to the binary, and the primary
and secondary black holes had dimensionless spin magnitudes of 0.67 and 0.12,
tilted at $45^\circ$ and $90^\circ$ to the orbital angular momentum. This
configuration produces a precession of the orbital plane which results in a
waveform that is not well approximated by a non-spinning signal. The analysis
was performed using the 15-dimensional parameterised model for a quasi-circular
black hole binary commonly used in LIGO-Virgo analyses (e.g.~\cite{PhysRevLett.116.241102,
O1BBH}), implemented in the LALInference package~\cite{Veitch:2015}.

\begin{figure}[]
	\centering 
	\includegraphics[scale=0.45,clip=true,angle=0]{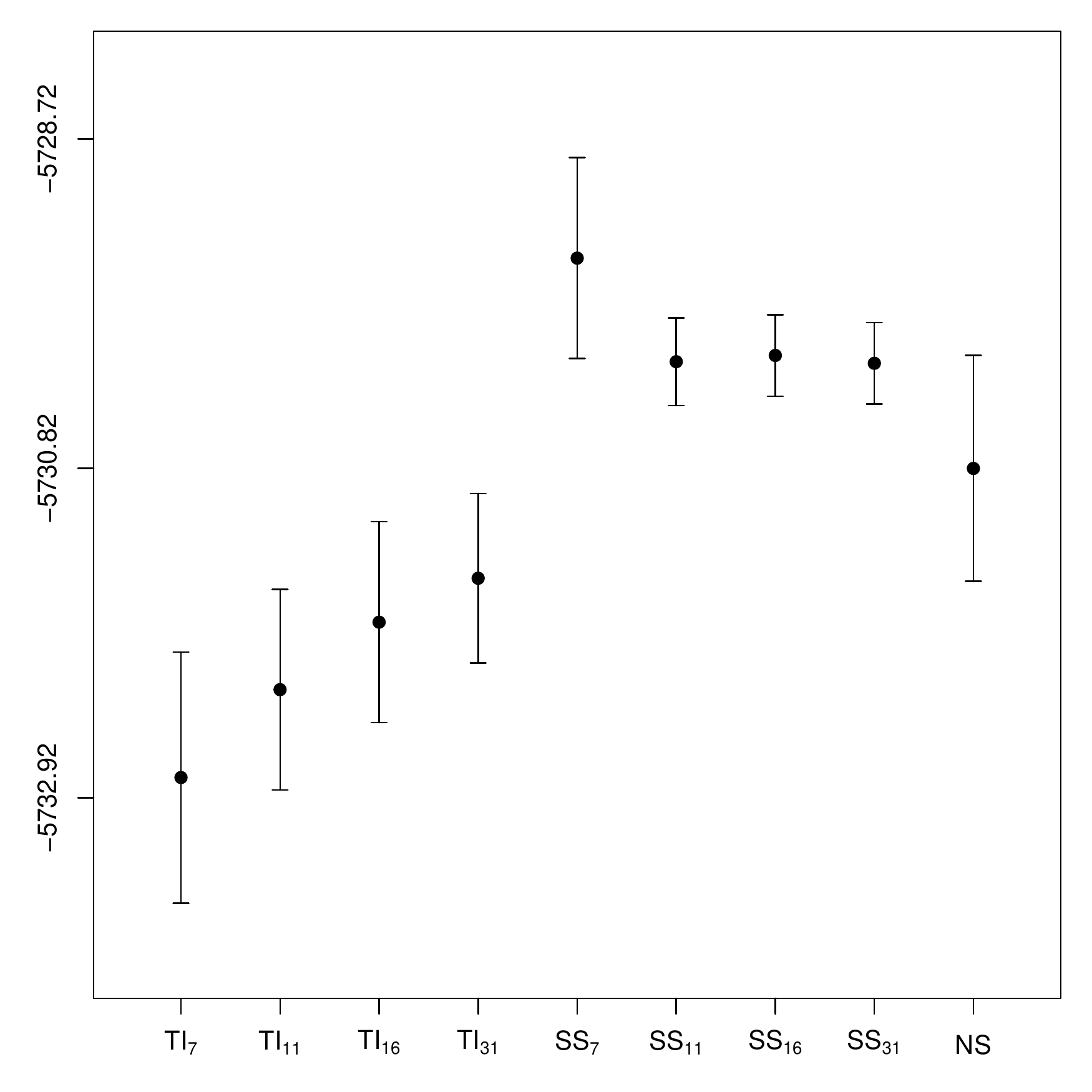}
	\caption{Evidence estimates $\pm$ 2 standard errors.  Subscripts in TI and SS stand for the number of temperatures.}
	\label{fig3}
\end{figure}

We estimate the marginal likelihood via NS, TI and SS.  For NS, we performed 32
runs with 2000 live points each. For TI and SS, we considered 31 temperatures, evenly spaced on a logarithmic scale, with 4700 samples from each.  From these simulations, we ran TI and SS to estimate the evidence for $K=7, 11, 16$, and $31$.  To compute the standard error of the evidence estimates, we applied the MBB method for different block lengths and took the one that yielded the maximum standard
deviation  as a conservative way of estimation. The results are displayed in Table~\ref{tab:results} and visualized in Figure~\ref{fig3}.
The evidence estimates of SS and NS are closer than of SS and TI. In the light of the performance of  TI in the simulation study, it seems that TI would have
needed more temperatures to achieve an evidence estimate consistent with both SS and NS. The standard error of the NS evidence estimate is quite large, especially
given that it was calculated for a large number of live points.  

\begin{table}
	\caption{\label{tab:results} Evidence estimates and corresponding standard errors, and the Bayes factor from the NS, TI and SS methods for different number of temperatures $K$.}
	\begin{ruledtabular}
		\begin{tabular}{ccccc}
			Method & $K$ & $\log z$ & SD& $\log B$  \\ \hline
			NS & - &-5730.82 & 0.36  & 103.33    \\ \hline
			TI   & 7 &-5732.79 &  0.40 & 101.36     \\
			      & 11 &-5732.23 & 0.32 &101.92\\
			      & 16 &-5731.80 & 0.32 &102.35\\
			      & 31 &-5731.52 & 0.27 &102.63\\ \hline
			SS & 7   &-5729.48 & 0.32 & 104.67  \\
			      & 11  &-5730.14 & 0.14 &104.01\\
			      & 16 &-5730.10 & 0.13 &104.05\\
			      & 31 &-5730.15 & 0.13 &104.00\\
		\end{tabular}
	\end{ruledtabular}
\end{table}

\section{Discussion}
SS is a method to estimate the marginal likelihood which has enjoyed great popularity in phylogenetics where it has been shown to work well.  It requires  less computational effort than TI to yield an accurate estimate of the evidence. In a simulation study with a simple Gaussian model, we have shown that it is less sensitive to the dispersal of the inverse temperature  $\beta$ values and achieves a higher accuracy with a smaller number of power posterior distributions.  To the best of our knowledge, it has not been applied for calculating the evidence of gravitational wave models yet.  Its implementation in this context should be straightforward since its main complexity lies with sampling from the power posterior, like TI.  However, this can be done by using the parallel tempering method, which has been widely implemented in GW software packages such as LALInference. 

The performance of SS depends on its specifications, such as the number of MCMC steps  $n$ in each parallel tempering chain, the number of temperatures $K$ and the distribution of the inverse temperature $\beta$ values.  In addition, it depends on how different the prior and the posterior are.  To mitigate the dependence on the prior distribution, we aim to explore a recent extension of SS  known as generalized steppingstone sampling \citep[GSS;][]{fan:2011}.  This method makes use of a reference distribution which aims to shorten the distance between the prior and the posterior.  Even though it requires posterior samples to construct the reference distribution, it could be more accurate than its simple version and require less steps to yield the same accuracy.  For this, the reference distribution needs to be a reasonable approximation of the posterior, otherwise it can dramatically fail \citep{Maturana:2017b}.  

One of the drawbacks of power posterior methods is the lack of a direct formula for the standard error of the evidence estimate.  In practice, the methods are run multiple times in order to obtain an empirical standard error estimate. This brute force approach  will prove too computationally expensive in most practial applications.  Alternatively,  the standard bootstrap has been applied.
It is computationally much cheaper than the brute force approach, but it does not take  the dependencies within and between the Markov chains  into account.  In this paper, we have proposed a moving block bootstrap method.  This approach has the ability to  allow for potential autocorrelation within the chains and cross-correlation between chains.  We showed in  Example~\ref{subsec:SDest} of our simulation study that the standard bootstrap severely underestimates the standard error in the presence of autocorrelation in  Markov chains but that the moving block bootstrap  significantly improves the standard error estimates of the evidence.  


\begin{acknowledgements}
  We thank Claudia Kirch for helpful discussions on the block-bootstrap for time series.  We also thank the New Zealand eScience Infrastructure (NeSI) for their high performance computing facilities, and the Centre for eResearch at the University of Auckland for their technical support. The Observatoire de la C\^{o}te d'Azur also provided support for this research. PM's and RM's work is supported by Grant 3714568 from the University of Auckland Faculty Research Development Fund and the DFG Grant KI 1443/3-1. JV is supported by STFC grant ST/K005014/1. NC's work is supported by NSF grants PHY-1806990 and PHY-1505373. This paper has been given LIGO Document Number P1800299.  All analysis was conducted in \textsf{R}, an open-source statistical software available on \textsf{CRAN} (cran.r-project.org) and LALInference.
\end{acknowledgements}
\pagebreak


\bibliographystyle{apsrev4-1}

\bibliography{PRDSS}

\begin{thebibliography}{52}%
\makeatletter
\providecommand \@ifxundefined [1]{%
 \@ifx{#1\undefined}
}%
\providecommand \@ifnum [1]{%
 \ifnum #1\expandafter \@firstoftwo
 \else \expandafter \@secondoftwo
 \fi
}%
\providecommand \@ifx [1]{%
 \ifx #1\expandafter \@firstoftwo
 \else \expandafter \@secondoftwo
 \fi
}%
\providecommand \natexlab [1]{#1}%
\providecommand \enquote  [1]{``#1''}%
\providecommand \bibnamefont  [1]{#1}%
\providecommand \bibfnamefont [1]{#1}%
\providecommand \citenamefont [1]{#1}%
\providecommand \href@noop [0]{\@secondoftwo}%
\providecommand \href [0]{\begingroup \@sanitize@url \@href}%
\providecommand \@href[1]{\@@startlink{#1}\@@href}%
\providecommand \@@href[1]{\endgroup#1\@@endlink}%
\providecommand \@sanitize@url [0]{\catcode `\\12\catcode `\$12\catcode
  `\&12\catcode `\#12\catcode `\^12\catcode `\_12\catcode `\%12\relax}%
\providecommand \@@startlink[1]{}%
\providecommand \@@endlink[0]{}%
\providecommand \url  [0]{\begingroup\@sanitize@url \@url }%
\providecommand \@url [1]{\endgroup\@href {#1}{\urlprefix }}%
\providecommand \urlprefix  [0]{URL }%
\providecommand \Eprint [0]{\href }%
\providecommand \doibase [0]{http://dx.doi.org/}%
\providecommand \selectlanguage [0]{\@gobble}%
\providecommand \bibinfo  [0]{\@secondoftwo}%
\providecommand \bibfield  [0]{\@secondoftwo}%
\providecommand \translation [1]{[#1]}%
\providecommand \BibitemOpen [0]{}%
\providecommand \bibitemStop [0]{}%
\providecommand \bibitemNoStop [0]{.\EOS\space}%
\providecommand \EOS [0]{\spacefactor3000\relax}%
\providecommand \BibitemShut  [1]{\csname bibitem#1\endcsname}%
\let\auto@bib@innerbib\@empty
\bibitem [{\citenamefont {{Saha}}\ and\ \citenamefont
  {{Williams}}(1994)}]{1994AJ....107.1295S}%
  \BibitemOpen
  \bibfield  {author} {\bibinfo {author} {\bibfnamefont {P.}~\bibnamefont
  {{Saha}}}\ and\ \bibinfo {author} {\bibfnamefont {T.~B.}\ \bibnamefont
  {{Williams}}},\ }\href {\doibase 10.1086/116942} {\bibfield  {journal}
  {\bibinfo  {journal} {The Astronomical Journal}\ }\textbf {\bibinfo {volume}
  {107}},\ \bibinfo {pages} {1295} (\bibinfo {year} {1994})}\BibitemShut
  {NoStop}%
\bibitem [{\citenamefont {Christensen}\ and\ \citenamefont
  {Meyer}(1998)}]{PhysRevD.58.082001}%
  \BibitemOpen
  \bibfield  {author} {\bibinfo {author} {\bibfnamefont {N.}~\bibnamefont
  {Christensen}}\ and\ \bibinfo {author} {\bibfnamefont {R.}~\bibnamefont
  {Meyer}},\ }\href {\doibase 10.1103/PhysRevD.58.082001} {\bibfield  {journal}
  {\bibinfo  {journal} {Phys. Rev. D}\ }\textbf {\bibinfo {volume} {58}},\
  \bibinfo {pages} {082001} (\bibinfo {year} {1998})}\BibitemShut {NoStop}%
\bibitem [{\citenamefont {Christensen}\ \emph {et~al.}(2001)\citenamefont
  {Christensen}, \citenamefont {Meyer}, \citenamefont {Knox},\ and\
  \citenamefont {Luey}}]{0264-9381-18-14-306}%
  \BibitemOpen
  \bibfield  {author} {\bibinfo {author} {\bibfnamefont {N.}~\bibnamefont
  {Christensen}}, \bibinfo {author} {\bibfnamefont {R.}~\bibnamefont {Meyer}},
  \bibinfo {author} {\bibfnamefont {L.}~\bibnamefont {Knox}}, \ and\ \bibinfo
  {author} {\bibfnamefont {B.}~\bibnamefont {Luey}},\ }\href
  {http://stacks.iop.org/0264-9381/18/i=14/a=306} {\bibfield  {journal}
  {\bibinfo  {journal} {Classical and Quantum Gravity}\ }\textbf {\bibinfo
  {volume} {18}},\ \bibinfo {pages} {2677} (\bibinfo {year}
  {2001})}\BibitemShut {NoStop}%
\bibitem [{\citenamefont {Knox}\ \emph {et~al.}(2001)\citenamefont {Knox},
  \citenamefont {Christensen},\ and\ \citenamefont
  {Skordis}}]{1538-4357-563-2-L95}%
  \BibitemOpen
  \bibfield  {author} {\bibinfo {author} {\bibfnamefont {L.}~\bibnamefont
  {Knox}}, \bibinfo {author} {\bibfnamefont {N.}~\bibnamefont {Christensen}}, \
  and\ \bibinfo {author} {\bibfnamefont {C.}~\bibnamefont {Skordis}},\ }\href
  {http://stacks.iop.org/1538-4357/563/i=2/a=L95} {\bibfield  {journal}
  {\bibinfo  {journal} {The Astrophysical Journal Letters}\ }\textbf {\bibinfo
  {volume} {563}},\ \bibinfo {pages} {L95} (\bibinfo {year}
  {2001})}\BibitemShut {NoStop}%
\bibitem [{\citenamefont
  {Sharma}(2017)}]{doi:10.1146/annurev-astro-082214-122339}%
  \BibitemOpen
  \bibfield  {author} {\bibinfo {author} {\bibfnamefont {S.}~\bibnamefont
  {Sharma}},\ }\href {\doibase 10.1146/annurev-astro-082214-122339} {\bibfield
  {journal} {\bibinfo  {journal} {Annual Review of Astronomy and Astrophysics}\
  }\textbf {\bibinfo {volume} {55}},\ \bibinfo {pages} {213} (\bibinfo {year}
  {2017})}\BibitemShut {NoStop}%
\bibitem [{\citenamefont {Hinshaw}\ \emph {et~al.}(2013)\citenamefont {Hinshaw}
  \emph {et~al.}}]{0067-0049-208-2-19}%
  \BibitemOpen
  \bibfield  {author} {\bibinfo {author} {\bibfnamefont {G.}~\bibnamefont
  {Hinshaw}} \emph {et~al.},\ }\href
  {http://stacks.iop.org/0067-0049/208/i=2/a=19} {\bibfield  {journal}
  {\bibinfo  {journal} {The Astrophysical Journal Supplement Series}\ }\textbf
  {\bibinfo {volume} {208}},\ \bibinfo {pages} {19} (\bibinfo {year}
  {2013})}\BibitemShut {NoStop}%
\bibitem [{\citenamefont {{Planck Collaboration}}\ \emph
  {et~al.}(2014)\citenamefont {{Planck Collaboration}}, \citenamefont {{Ade, P.
  A. R.}} \emph {et~al.}}]{Planck2013}%
  \BibitemOpen
  \bibfield  {author} {\bibinfo {author} {\bibnamefont {{Planck
  Collaboration}}}, \bibinfo {author} {\bibnamefont {{Ade, P. A. R.}}},  \emph
  {et~al.},\ }\href {\doibase 10.1051/0004-6361/201321591} {\bibfield
  {journal} {\bibinfo  {journal} {Astronomy \& Astrophysics}\ }\textbf
  {\bibinfo {volume} {571}},\ \bibinfo {pages} {A16} (\bibinfo {year}
  {2014})}\BibitemShut {NoStop}%
\bibitem [{\citenamefont {{Planck Collaboration}}\ \emph
  {et~al.}(2016)\citenamefont {{Planck Collaboration}}, \citenamefont {{Ade, P.
  A. R.}} \emph {et~al.}}]{Planck2015}%
  \BibitemOpen
  \bibfield  {author} {\bibinfo {author} {\bibnamefont {{Planck
  Collaboration}}}, \bibinfo {author} {\bibnamefont {{Ade, P. A. R.}}},  \emph
  {et~al.},\ }\href {\doibase 10.1051/0004-6361/201525830} {\bibfield
  {journal} {\bibinfo  {journal} {Astronomy \& Astrophysics}\ }\textbf
  {\bibinfo {volume} {594}},\ \bibinfo {pages} {A13} (\bibinfo {year}
  {2016})}\BibitemShut {NoStop}%
\bibitem [{\citenamefont {Abbott}\ \emph
  {et~al.}(2016{\natexlab{a}})\citenamefont {Abbott} \emph
  {et~al.}}]{PhysRevLett.116.061102}%
  \BibitemOpen
  \bibfield  {author} {\bibinfo {author} {\bibfnamefont {B.~P.}\ \bibnamefont
  {Abbott}} \emph {et~al.} (\bibinfo {collaboration} {LIGO Scientific
  Collaboration and Virgo Collaboration}),\ }\href {\doibase
  10.1103/PhysRevLett.116.061102} {\bibfield  {journal} {\bibinfo  {journal}
  {Phys. Rev. Lett.}\ }\textbf {\bibinfo {volume} {116}},\ \bibinfo {pages}
  {061102} (\bibinfo {year} {2016}{\natexlab{a}})}\BibitemShut {NoStop}%
\bibitem [{\citenamefont {Abbott}\ \emph
  {et~al.}(2016{\natexlab{b}})\citenamefont {Abbott} \emph
  {et~al.}}]{PhysRevLett.116.241102}%
  \BibitemOpen
  \bibfield  {author} {\bibinfo {author} {\bibfnamefont {B.~P.}\ \bibnamefont
  {Abbott}} \emph {et~al.} (\bibinfo {collaboration} {LIGO Scientific
  Collaboration and Virgo Collaboration}),\ }\href {\doibase
  10.1103/PhysRevLett.116.241102} {\bibfield  {journal} {\bibinfo  {journal}
  {Phys. Rev. Lett.}\ }\textbf {\bibinfo {volume} {116}},\ \bibinfo {pages}
  {241102} (\bibinfo {year} {2016}{\natexlab{b}})}\BibitemShut {NoStop}%
\bibitem [{\citenamefont {Meyer}\ and\ \citenamefont
  {Christensen}()}]{doi:10.1111/j.1740-9713.2016.00896.x}%
  \BibitemOpen
  \bibfield  {author} {\bibinfo {author} {\bibfnamefont {R.}~\bibnamefont
  {Meyer}}\ and\ \bibinfo {author} {\bibfnamefont {N.}~\bibnamefont
  {Christensen}},\ }\href {\doibase 10.1111/j.1740-9713.2016.00896.x}
  {\bibfield  {journal} {\bibinfo  {journal} {Significance}\ }\textbf {\bibinfo
  {volume} {13}},\ \bibinfo {pages} {20}}\BibitemShut {NoStop}%
\bibitem [{\citenamefont {Abbott}\ \emph
  {et~al.}(2017{\natexlab{a}})\citenamefont {Abbott} \emph
  {et~al.}}]{PhysRevLett.119.161101}%
  \BibitemOpen
  \bibfield  {author} {\bibinfo {author} {\bibfnamefont {B.~P.}\ \bibnamefont
  {Abbott}} \emph {et~al.} (\bibinfo {collaboration} {LIGO Scientific
  Collaboration and Virgo Collaboration}),\ }\href {\doibase
  10.1103/PhysRevLett.119.161101} {\bibfield  {journal} {\bibinfo  {journal}
  {Phys. Rev. Lett.}\ }\textbf {\bibinfo {volume} {119}},\ \bibinfo {pages}
  {161101} (\bibinfo {year} {2017}{\natexlab{a}})}\BibitemShut {NoStop}%
\bibitem [{\citenamefont {Zhu}\ \emph {et~al.}(2017)\citenamefont {Zhu},
  \citenamefont {Chen}, \citenamefont {Hu},\ and\ \citenamefont
  {Zhang}}]{doi:10.1093/nsr/nwx044}%
  \BibitemOpen
  \bibfield  {author} {\bibinfo {author} {\bibfnamefont {J.}~\bibnamefont
  {Zhu}}, \bibinfo {author} {\bibfnamefont {J.}~\bibnamefont {Chen}}, \bibinfo
  {author} {\bibfnamefont {W.}~\bibnamefont {Hu}}, \ and\ \bibinfo {author}
  {\bibfnamefont {B.}~\bibnamefont {Zhang}},\ }\href {\doibase
  10.1093/nsr/nwx044} {\bibfield  {journal} {\bibinfo  {journal} {National
  Science Review}\ }\textbf {\bibinfo {volume} {4}},\ \bibinfo {pages} {627}
  (\bibinfo {year} {2017})}\BibitemShut {NoStop}%
\bibitem [{\citenamefont {Goldstein}\ \emph {et~al.}(2017)\citenamefont
  {Goldstein}, \citenamefont {Veres}, \citenamefont {Burns}, \citenamefont
  {Briggs}, \citenamefont {Hamburg}, \citenamefont {Kocevski}, \citenamefont
  {Wilson-Hodge}, \citenamefont {Preece}, \citenamefont {Poolakkil},
  \citenamefont {Roberts}, \citenamefont {Hui}, \citenamefont {Connaughton},
  \citenamefont {Racusin}, \citenamefont {von Kienlin}, \citenamefont {Canton},
  \citenamefont {Christensen}, \citenamefont {Littenberg}, \citenamefont
  {Siellez}, \citenamefont {Blackburn}, \citenamefont {Broida}, \citenamefont
  {Bissaldi}, \citenamefont {Cleveland}, \citenamefont {Gibby}, \citenamefont
  {Giles}, \citenamefont {Kippen}, \citenamefont {McBreen}, \citenamefont
  {McEnery}, \citenamefont {Meegan}, \citenamefont {Paciesas},\ and\
  \citenamefont {Stanbro}}]{2041-8205-848-2-L14}%
  \BibitemOpen
  \bibfield  {author} {\bibinfo {author} {\bibfnamefont {A.}~\bibnamefont
  {Goldstein}}, \bibinfo {author} {\bibfnamefont {P.}~\bibnamefont {Veres}},
  \bibinfo {author} {\bibfnamefont {E.}~\bibnamefont {Burns}}, \bibinfo
  {author} {\bibfnamefont {M.~S.}\ \bibnamefont {Briggs}}, \bibinfo {author}
  {\bibfnamefont {R.}~\bibnamefont {Hamburg}}, \bibinfo {author} {\bibfnamefont
  {D.}~\bibnamefont {Kocevski}}, \bibinfo {author} {\bibfnamefont {C.~A.}\
  \bibnamefont {Wilson-Hodge}}, \bibinfo {author} {\bibfnamefont {R.~D.}\
  \bibnamefont {Preece}}, \bibinfo {author} {\bibfnamefont {S.}~\bibnamefont
  {Poolakkil}}, \bibinfo {author} {\bibfnamefont {O.~J.}\ \bibnamefont
  {Roberts}}, \bibinfo {author} {\bibfnamefont {C.~M.}\ \bibnamefont {Hui}},
  \bibinfo {author} {\bibfnamefont {V.}~\bibnamefont {Connaughton}}, \bibinfo
  {author} {\bibfnamefont {J.}~\bibnamefont {Racusin}}, \bibinfo {author}
  {\bibfnamefont {A.}~\bibnamefont {von Kienlin}}, \bibinfo {author}
  {\bibfnamefont {T.~D.}\ \bibnamefont {Canton}}, \bibinfo {author}
  {\bibfnamefont {N.}~\bibnamefont {Christensen}}, \bibinfo {author}
  {\bibfnamefont {T.}~\bibnamefont {Littenberg}}, \bibinfo {author}
  {\bibfnamefont {K.}~\bibnamefont {Siellez}}, \bibinfo {author} {\bibfnamefont
  {L.}~\bibnamefont {Blackburn}}, \bibinfo {author} {\bibfnamefont
  {J.}~\bibnamefont {Broida}}, \bibinfo {author} {\bibfnamefont
  {E.}~\bibnamefont {Bissaldi}}, \bibinfo {author} {\bibfnamefont {W.~H.}\
  \bibnamefont {Cleveland}}, \bibinfo {author} {\bibfnamefont {M.~H.}\
  \bibnamefont {Gibby}}, \bibinfo {author} {\bibfnamefont {M.~M.}\ \bibnamefont
  {Giles}}, \bibinfo {author} {\bibfnamefont {R.~M.}\ \bibnamefont {Kippen}},
  \bibinfo {author} {\bibfnamefont {S.}~\bibnamefont {McBreen}}, \bibinfo
  {author} {\bibfnamefont {J.}~\bibnamefont {McEnery}}, \bibinfo {author}
  {\bibfnamefont {C.~A.}\ \bibnamefont {Meegan}}, \bibinfo {author}
  {\bibfnamefont {W.~S.}\ \bibnamefont {Paciesas}}, \ and\ \bibinfo {author}
  {\bibfnamefont {M.}~\bibnamefont {Stanbro}},\ }\href
  {http://stacks.iop.org/2041-8205/848/i=2/a=L14} {\bibfield  {journal}
  {\bibinfo  {journal} {The Astrophysical Journal Letters}\ }\textbf {\bibinfo
  {volume} {848}},\ \bibinfo {pages} {L14} (\bibinfo {year}
  {2017})}\BibitemShut {NoStop}%
\bibitem [{\citenamefont {Aasi}\ \emph {et~al.}(2015)\citenamefont {Aasi} \emph
  {et~al.}}]{0264-9381-32-7-074001}%
  \BibitemOpen
  \bibfield  {author} {\bibinfo {author} {\bibfnamefont {J.}~\bibnamefont
  {Aasi}} \emph {et~al.} (\bibinfo {collaboration} {LIGO Scientific
  Collaboration and Virgo Collaboration}),\ }\href
  {http://stacks.iop.org/0264-9381/32/i=7/a=074001} {\bibfield  {journal}
  {\bibinfo  {journal} {Classical and Quantum Gravity}\ }\textbf {\bibinfo
  {volume} {32}},\ \bibinfo {pages} {074001} (\bibinfo {year}
  {2015})}\BibitemShut {NoStop}%
\bibitem [{\citenamefont {Acernese}\ \emph {et~al.}(2015)\citenamefont
  {Acernese} \emph {et~al.}}]{0264-9381-32-2-024001}%
  \BibitemOpen
  \bibfield  {author} {\bibinfo {author} {\bibfnamefont {F.}~\bibnamefont
  {Acernese}} \emph {et~al.},\ }\href
  {http://stacks.iop.org/0264-9381/32/i=2/a=024001} {\bibfield  {journal}
  {\bibinfo  {journal} {Classical and Quantum Gravity}\ }\textbf {\bibinfo
  {volume} {32}},\ \bibinfo {pages} {024001} (\bibinfo {year}
  {2015})}\BibitemShut {NoStop}%
\bibitem [{\citenamefont {Abbott}\ \emph
  {et~al.}(2017{\natexlab{b}})\citenamefont {Abbott} \emph
  {et~al.}}]{2041-8205-848-2-L12}%
  \BibitemOpen
  \bibfield  {author} {\bibinfo {author} {\bibfnamefont {B.~P.}\ \bibnamefont
  {Abbott}} \emph {et~al.},\ }\href
  {http://stacks.iop.org/2041-8205/848/i=2/a=L12} {\bibfield  {journal}
  {\bibinfo  {journal} {The Astrophysical Journal Letters}\ }\textbf {\bibinfo
  {volume} {848}},\ \bibinfo {pages} {L12} (\bibinfo {year}
  {2017}{\natexlab{b}})}\BibitemShut {NoStop}%
\bibitem [{\citenamefont {Singer}\ and\ \citenamefont
  {Price}(2016)}]{PhysRevD.93.024013}%
  \BibitemOpen
  \bibfield  {author} {\bibinfo {author} {\bibfnamefont {L.~P.}\ \bibnamefont
  {Singer}}\ and\ \bibinfo {author} {\bibfnamefont {L.~R.}\ \bibnamefont
  {Price}},\ }\href {\doibase 10.1103/PhysRevD.93.024013} {\bibfield  {journal}
  {\bibinfo  {journal} {Phys. Rev. D}\ }\textbf {\bibinfo {volume} {93}},\
  \bibinfo {pages} {024013} (\bibinfo {year} {2016})}\BibitemShut {NoStop}%
\bibitem [{\citenamefont {Veitch}\ \emph {et~al.}(2015)\citenamefont {Veitch},
  \citenamefont {Raymond}, \citenamefont {Farr}, \citenamefont {Farr},
  \citenamefont {Graff}, \citenamefont {Vitale}, \citenamefont {Aylott},
  \citenamefont {Blackburn}, \citenamefont {Christensen}, \citenamefont
  {Coughlin}, \citenamefont {Del~Pozzo}, \citenamefont {Feroz}, \citenamefont
  {Gair}, \citenamefont {Haster}, \citenamefont {Kalogera}, \citenamefont
  {Littenberg}, \citenamefont {Mandel}, \citenamefont {O'Shaughnessy},
  \citenamefont {Pitkin}, \citenamefont {Rodriguez}, \citenamefont {R\"over},
  \citenamefont {Sidery}, \citenamefont {Smith}, \citenamefont {Van Der~Sluys},
  \citenamefont {Vecchio}, \citenamefont {Vousden},\ and\ \citenamefont
  {Wade}}]{Veitch:2015}%
  \BibitemOpen
  \bibfield  {author} {\bibinfo {author} {\bibfnamefont {J.}~\bibnamefont
  {Veitch}}, \bibinfo {author} {\bibfnamefont {V.}~\bibnamefont {Raymond}},
  \bibinfo {author} {\bibfnamefont {B.}~\bibnamefont {Farr}}, \bibinfo {author}
  {\bibfnamefont {W.}~\bibnamefont {Farr}}, \bibinfo {author} {\bibfnamefont
  {P.}~\bibnamefont {Graff}}, \bibinfo {author} {\bibfnamefont
  {S.}~\bibnamefont {Vitale}}, \bibinfo {author} {\bibfnamefont
  {B.}~\bibnamefont {Aylott}}, \bibinfo {author} {\bibfnamefont
  {K.}~\bibnamefont {Blackburn}}, \bibinfo {author} {\bibfnamefont
  {N.}~\bibnamefont {Christensen}}, \bibinfo {author} {\bibfnamefont
  {M.}~\bibnamefont {Coughlin}}, \bibinfo {author} {\bibfnamefont
  {W.}~\bibnamefont {Del~Pozzo}}, \bibinfo {author} {\bibfnamefont
  {F.}~\bibnamefont {Feroz}}, \bibinfo {author} {\bibfnamefont
  {J.}~\bibnamefont {Gair}}, \bibinfo {author} {\bibfnamefont {C.-J.}\
  \bibnamefont {Haster}}, \bibinfo {author} {\bibfnamefont {V.}~\bibnamefont
  {Kalogera}}, \bibinfo {author} {\bibfnamefont {T.}~\bibnamefont
  {Littenberg}}, \bibinfo {author} {\bibfnamefont {I.}~\bibnamefont {Mandel}},
  \bibinfo {author} {\bibfnamefont {R.}~\bibnamefont {O'Shaughnessy}}, \bibinfo
  {author} {\bibfnamefont {M.}~\bibnamefont {Pitkin}}, \bibinfo {author}
  {\bibfnamefont {C.}~\bibnamefont {Rodriguez}}, \bibinfo {author}
  {\bibfnamefont {C.}~\bibnamefont {R\"over}}, \bibinfo {author} {\bibfnamefont
  {T.}~\bibnamefont {Sidery}}, \bibinfo {author} {\bibfnamefont
  {R.}~\bibnamefont {Smith}}, \bibinfo {author} {\bibfnamefont
  {M.}~\bibnamefont {Van Der~Sluys}}, \bibinfo {author} {\bibfnamefont
  {A.}~\bibnamefont {Vecchio}}, \bibinfo {author} {\bibfnamefont
  {W.}~\bibnamefont {Vousden}}, \ and\ \bibinfo {author} {\bibfnamefont
  {L.}~\bibnamefont {Wade}},\ }\href {\doibase 10.1103/PhysRevD.91.042003}
  {\bibfield  {journal} {\bibinfo  {journal} {Phys. Rev. D}\ }\textbf {\bibinfo
  {volume} {91}},\ \bibinfo {pages} {042003} (\bibinfo {year}
  {2015})}\BibitemShut {NoStop}%
\bibitem [{\citenamefont {Abbott}\ \emph
  {et~al.}(2016{\natexlab{c}})\citenamefont {Abbott} \emph
  {et~al.}}]{PhysRevX.6.041014}%
  \BibitemOpen
  \bibfield  {author} {\bibinfo {author} {\bibfnamefont {B.~P.}\ \bibnamefont
  {Abbott}} \emph {et~al.} (\bibinfo {collaboration} {LIGO Scientific
  Collaboration and Virgo Collaboration}),\ }\href {\doibase
  10.1103/PhysRevX.6.041014} {\bibfield  {journal} {\bibinfo  {journal} {Phys.
  Rev. X}\ }\textbf {\bibinfo {volume} {6}},\ \bibinfo {pages} {041014}
  (\bibinfo {year} {2016}{\natexlab{c}})}\BibitemShut {NoStop}%
\bibitem [{\citenamefont {Abbott}\ \emph
  {et~al.}(2016{\natexlab{d}})\citenamefont {Abbott} \emph {et~al.}}]{O1BBH}%
  \BibitemOpen
  \bibfield  {author} {\bibinfo {author} {\bibfnamefont {B.~P.}\ \bibnamefont
  {Abbott}} \emph {et~al.} (\bibinfo {collaboration} {LIGO Scientific
  Collaboration, Virgo Collaboration}),\ }\href {\doibase
  10.1103/PhysRevX.6.041015} {\bibfield  {journal} {\bibinfo  {journal} {Phys.
  Rev. X}\ }\textbf {\bibinfo {volume} {6}},\ \bibinfo {pages} {041015}
  (\bibinfo {year} {2016}{\natexlab{d}})},\ \Eprint
  {http://arxiv.org/abs/1606.04856} {arXiv:1606.04856 [gr-qc]} \BibitemShut
  {NoStop}%
\bibitem [{\citenamefont {Abbott}\ \emph
  {et~al.}(2018{\natexlab{a}})\citenamefont {Abbott} \emph
  {et~al.}}]{Abbott:2018wiz}%
  \BibitemOpen
  \bibfield  {author} {\bibinfo {author} {\bibfnamefont {B.~P.}\ \bibnamefont
  {Abbott}} \emph {et~al.} (\bibinfo {collaboration} {Virgo, LIGO
  Scientific}),\ }\href@noop {} {\  (\bibinfo {year} {2018}{\natexlab{a}})},\
  \Eprint {http://arxiv.org/abs/1805.11579} {arXiv:1805.11579 [gr-qc]}
  \BibitemShut {NoStop}%
\bibitem [{\citenamefont {Abbott}\ \emph
  {et~al.}(2016{\natexlab{e}})\citenamefont {Abbott} \emph
  {et~al.}}]{PhysRevLett.116.221101}%
  \BibitemOpen
  \bibfield  {author} {\bibinfo {author} {\bibfnamefont {B.~P.}\ \bibnamefont
  {Abbott}} \emph {et~al.} (\bibinfo {collaboration} {LIGO Scientific and Virgo
  Collaborations}),\ }\href {\doibase 10.1103/PhysRevLett.116.221101}
  {\bibfield  {journal} {\bibinfo  {journal} {Phys. Rev. Lett.}\ }\textbf
  {\bibinfo {volume} {116}},\ \bibinfo {pages} {221101} (\bibinfo {year}
  {2016}{\natexlab{e}})}\BibitemShut {NoStop}%
\bibitem [{\citenamefont {Abbott}\ \emph
  {et~al.}(2018{\natexlab{b}})\citenamefont {Abbott} \emph
  {et~al.}}]{Abbott:2018exr}%
  \BibitemOpen
  \bibfield  {author} {\bibinfo {author} {\bibfnamefont {B.~P.}\ \bibnamefont
  {Abbott}} \emph {et~al.} (\bibinfo {collaboration} {Virgo, LIGO
  Scientific}),\ }\href@noop {} {\  (\bibinfo {year} {2018}{\natexlab{b}})},\
  \Eprint {http://arxiv.org/abs/1805.11581} {arXiv:1805.11581 [gr-qc]}
  \BibitemShut {NoStop}%
\bibitem [{\citenamefont {Weinberg}(2018)}]{Weinberg:2018icl}%
  \BibitemOpen
  \bibfield  {author} {\bibinfo {author} {\bibfnamefont {N.~N.}\ \bibnamefont
  {Weinberg}} (\bibinfo {collaboration} {Virgo, LIGO Scientific}),\ }\href@noop
  {} {\  (\bibinfo {year} {2018})},\ \Eprint {http://arxiv.org/abs/1808.08676}
  {arXiv:1808.08676 [astro-ph.HE]} \BibitemShut {NoStop}%
\bibitem [{\citenamefont {Smith}\ and\ \citenamefont
  {Thrane}(2018)}]{PhysRevX.8.021019}%
  \BibitemOpen
  \bibfield  {author} {\bibinfo {author} {\bibfnamefont {R.}~\bibnamefont
  {Smith}}\ and\ \bibinfo {author} {\bibfnamefont {E.}~\bibnamefont {Thrane}},\
  }\href {\doibase 10.1103/PhysRevX.8.021019} {\bibfield  {journal} {\bibinfo
  {journal} {Phys. Rev. X}\ }\textbf {\bibinfo {volume} {8}},\ \bibinfo {pages}
  {021019} (\bibinfo {year} {2018})}\BibitemShut {NoStop}%
\bibitem [{\citenamefont {Abbott}\ \emph
  {et~al.}(2017{\natexlab{c}})\citenamefont {Abbott} \emph
  {et~al.}}]{PhysRevLett.119.141101}%
  \BibitemOpen
  \bibfield  {author} {\bibinfo {author} {\bibfnamefont {B.~P.}\ \bibnamefont
  {Abbott}} \emph {et~al.} (\bibinfo {collaboration} {LIGO Scientific
  Collaboration and Virgo Collaboration}),\ }\href {\doibase
  10.1103/PhysRevLett.119.141101} {\bibfield  {journal} {\bibinfo  {journal}
  {Phys. Rev. Lett.}\ }\textbf {\bibinfo {volume} {119}},\ \bibinfo {pages}
  {141101} (\bibinfo {year} {2017}{\natexlab{c}})}\BibitemShut {NoStop}%
\bibitem [{\citenamefont {Callister}\ \emph {et~al.}(2017)\citenamefont
  {Callister}, \citenamefont {Biscoveanu}, \citenamefont {Christensen},
  \citenamefont {Isi}, \citenamefont {Matas}, \citenamefont {Minazzoli},
  \citenamefont {Regimbau}, \citenamefont {Sakellariadou}, \citenamefont
  {Tasson},\ and\ \citenamefont {Thrane}}]{PhysRevX.7.041058}%
  \BibitemOpen
  \bibfield  {author} {\bibinfo {author} {\bibfnamefont {T.}~\bibnamefont
  {Callister}}, \bibinfo {author} {\bibfnamefont {A.~S.}\ \bibnamefont
  {Biscoveanu}}, \bibinfo {author} {\bibfnamefont {N.}~\bibnamefont
  {Christensen}}, \bibinfo {author} {\bibfnamefont {M.}~\bibnamefont {Isi}},
  \bibinfo {author} {\bibfnamefont {A.}~\bibnamefont {Matas}}, \bibinfo
  {author} {\bibfnamefont {O.}~\bibnamefont {Minazzoli}}, \bibinfo {author}
  {\bibfnamefont {T.}~\bibnamefont {Regimbau}}, \bibinfo {author}
  {\bibfnamefont {M.}~\bibnamefont {Sakellariadou}}, \bibinfo {author}
  {\bibfnamefont {J.}~\bibnamefont {Tasson}}, \ and\ \bibinfo {author}
  {\bibfnamefont {E.}~\bibnamefont {Thrane}},\ }\href {\doibase
  10.1103/PhysRevX.7.041058} {\bibfield  {journal} {\bibinfo  {journal} {Phys.
  Rev. X}\ }\textbf {\bibinfo {volume} {7}},\ \bibinfo {pages} {041058}
  (\bibinfo {year} {2017})}\BibitemShut {NoStop}%
\bibitem [{\citenamefont {Abbott}\ \emph
  {et~al.}(2018{\natexlab{c}})\citenamefont {Abbott} \emph
  {et~al.}}]{PhysRevLett.120.201102}%
  \BibitemOpen
  \bibfield  {author} {\bibinfo {author} {\bibfnamefont {B.~P.}\ \bibnamefont
  {Abbott}} \emph {et~al.} (\bibinfo {collaboration} {LIGO Scientific
  Collaboration and Virgo Collaboration}),\ }\href {\doibase
  10.1103/PhysRevLett.120.201102} {\bibfield  {journal} {\bibinfo  {journal}
  {Phys. Rev. Lett.}\ }\textbf {\bibinfo {volume} {120}},\ \bibinfo {pages}
  {201102} (\bibinfo {year} {2018}{\natexlab{c}})}\BibitemShut {NoStop}%
\bibitem [{\citenamefont {Newton}\ and\ \citenamefont
  {Raftery}(1994)}]{Newton:Raftery:1994}%
  \BibitemOpen
  \bibfield  {author} {\bibinfo {author} {\bibfnamefont {M.~A.}\ \bibnamefont
  {Newton}}\ and\ \bibinfo {author} {\bibfnamefont {A.~E.}\ \bibnamefont
  {Raftery}},\ }\href@noop {} {\bibfield  {journal} {\bibinfo  {journal} {J.
  Roy. Statist. Soc. Ser. B}\ }\textbf {\bibinfo {volume} {56}},\ \bibinfo
  {pages} {3} (\bibinfo {year} {1994})}\BibitemShut {NoStop}%
\bibitem [{\citenamefont {Xie}\ \emph {et~al.}(2011)\citenamefont {Xie},
  \citenamefont {Lewis}, \citenamefont {Fan}, \citenamefont {Kuo},\ and\
  \citenamefont {Chen}}]{Xie:Lewis:Fan:Kuo:Chen:2011}%
  \BibitemOpen
  \bibfield  {author} {\bibinfo {author} {\bibfnamefont {W.}~\bibnamefont
  {Xie}}, \bibinfo {author} {\bibfnamefont {P.~O.}\ \bibnamefont {Lewis}},
  \bibinfo {author} {\bibfnamefont {Y.}~\bibnamefont {Fan}}, \bibinfo {author}
  {\bibfnamefont {L.}~\bibnamefont {Kuo}}, \ and\ \bibinfo {author}
  {\bibfnamefont {M.-H.}\ \bibnamefont {Chen}},\ }\href {\doibase
  10.1093/sysbio/syq085} {\bibfield  {journal} {\bibinfo  {journal} {Syst.
  Biol.}\ }\textbf {\bibinfo {volume} {60}},\ \bibinfo {pages} {150} (\bibinfo
  {year} {2011})}\BibitemShut {NoStop}%
\bibitem [{\citenamefont {Lartillot}\ and\ \citenamefont
  {Philippe}(2006)}]{Lartillot:Philippe:2006}%
  \BibitemOpen
  \bibfield  {author} {\bibinfo {author} {\bibfnamefont {N.}~\bibnamefont
  {Lartillot}}\ and\ \bibinfo {author} {\bibfnamefont {H.}~\bibnamefont
  {Philippe}},\ }\href {\doibase 10.1080/10635150500433722} {\bibfield
  {journal} {\bibinfo  {journal} {Syst. Biol.}\ }\textbf {\bibinfo {volume}
  {55}},\ \bibinfo {pages} {195} (\bibinfo {year} {2006})}\BibitemShut
  {NoStop}%
\bibitem [{\citenamefont {Friel}\ and\ \citenamefont
  {Pettitt}(2008)}]{Friel:2008}%
  \BibitemOpen
  \bibfield  {author} {\bibinfo {author} {\bibfnamefont {N.}~\bibnamefont
  {Friel}}\ and\ \bibinfo {author} {\bibfnamefont {A.~N.}\ \bibnamefont
  {Pettitt}},\ }\href {\doibase 10.1111/j.1467-9868.2007.00650.x} {\bibfield
  {journal} {\bibinfo  {journal} {J. Roy. Stat. Soc. B}\ }\textbf {\bibinfo
  {volume} {70}},\ \bibinfo {pages} {589} (\bibinfo {year} {2008})}\BibitemShut
  {NoStop}%
\bibitem [{\citenamefont {Neal}(2001)}]{Neal:2001}%
  \BibitemOpen
  \bibfield  {author} {\bibinfo {author} {\bibfnamefont {R.~M.}\ \bibnamefont
  {Neal}},\ }\href {\doibase 10.1023/A:1008923215028} {\bibfield  {journal}
  {\bibinfo  {journal} {Stat. Comput.}\ }\textbf {\bibinfo {volume} {11}},\
  \bibinfo {pages} {125} (\bibinfo {year} {2001})}\BibitemShut {NoStop}%
\bibitem [{\citenamefont {Skilling}(2006)}]{Skilling:2006}%
  \BibitemOpen
  \bibfield  {author} {\bibinfo {author} {\bibfnamefont {J.}~\bibnamefont
  {Skilling}},\ }\href@noop {} {\bibfield  {journal} {\bibinfo  {journal}
  {Bayesian Analysis}\ }\textbf {\bibinfo {volume} {1}},\ \bibinfo {pages}
  {833} (\bibinfo {year} {2006})}\BibitemShut {NoStop}%
\bibitem [{\citenamefont {Veitch}\ and\ \citenamefont
  {Vecchio}(2010)}]{Veitch:2010}%
  \BibitemOpen
  \bibfield  {author} {\bibinfo {author} {\bibfnamefont {J.}~\bibnamefont
  {Veitch}}\ and\ \bibinfo {author} {\bibfnamefont {A.}~\bibnamefont
  {Vecchio}},\ }\href {\doibase 10.1103/PhysRevD.81.062003} {\bibfield
  {journal} {\bibinfo  {journal} {Phys. Rev. D}\ }\textbf {\bibinfo {volume}
  {81}},\ \bibinfo {pages} {062003} (\bibinfo {year} {2010})}\BibitemShut
  {NoStop}%
\bibitem [{\citenamefont {{Brewer}}\ and\ \citenamefont
  {{Donovan}}(2015)}]{Brewer:Donovan:2015}%
  \BibitemOpen
  \bibfield  {author} {\bibinfo {author} {\bibfnamefont {B.~J.}\ \bibnamefont
  {{Brewer}}}\ and\ \bibinfo {author} {\bibfnamefont {C.~P.}\ \bibnamefont
  {{Donovan}}},\ }\href {\doibase 10.1093/mnras/stv199} {\bibfield  {journal}
  {\bibinfo  {journal} {Mon. Not. R. Astron. Soc.}\ }\textbf {\bibinfo {volume}
  {448}},\ \bibinfo {pages} {3206} (\bibinfo {year} {2015})}\BibitemShut
  {NoStop}%
\bibitem [{\citenamefont {Feroz}\ \emph {et~al.}(2009)\citenamefont {Feroz},
  \citenamefont {Hobson},\ and\ \citenamefont {Bridges}}]{Feroz:2009}%
  \BibitemOpen
  \bibfield  {author} {\bibinfo {author} {\bibfnamefont {F.}~\bibnamefont
  {Feroz}}, \bibinfo {author} {\bibfnamefont {M.}~\bibnamefont {Hobson}}, \
  and\ \bibinfo {author} {\bibfnamefont {M.}~\bibnamefont {Bridges}},\ }\href
  {\doibase 10.1111/j.1365-2966.2009.14548.x} {\bibfield  {journal} {\bibinfo
  {journal} {Monthly Notices of the Royal Astronomical Society}\ }\textbf
  {\bibinfo {volume} {398}},\ \bibinfo {pages} {1601} (\bibinfo {year}
  {2009})},\ \bibinfo {note} {cited By 723}\BibitemShut {NoStop}%
\bibitem [{\citenamefont {Henderson}\ \emph {et~al.}(2017)\citenamefont
  {Henderson}, \citenamefont {Goggans},\ and\ \citenamefont
  {Cao}}]{Henderson:2017}%
  \BibitemOpen
  \bibfield  {author} {\bibinfo {author} {\bibfnamefont {R.}~\bibnamefont
  {Henderson}}, \bibinfo {author} {\bibfnamefont {P.}~\bibnamefont {Goggans}},
  \ and\ \bibinfo {author} {\bibfnamefont {L.}~\bibnamefont {Cao}},\ }\href
  {\doibase 10.1016/j.dsp.2017.07.021} {\bibfield  {journal} {\bibinfo
  {journal} {Digital Signal Processing: A Review Journal}\ }\textbf {\bibinfo
  {volume} {70}},\ \bibinfo {pages} {84} (\bibinfo {year} {2017})},\ \bibinfo
  {note} {cited By 1}\BibitemShut {NoStop}%
\bibitem [{\citenamefont {Maturana~Russel}\ \emph {et~al.}(2018)\citenamefont
  {Maturana~Russel}, \citenamefont {Brewer}, \citenamefont {Klaere},\ and\
  \citenamefont {Bouckaert}}]{Maturana:2017b}%
  \BibitemOpen
  \bibfield  {author} {\bibinfo {author} {\bibfnamefont {P.}~\bibnamefont
  {Maturana~Russel}}, \bibinfo {author} {\bibfnamefont {B.~J.}\ \bibnamefont
  {Brewer}}, \bibinfo {author} {\bibfnamefont {S.}~\bibnamefont {Klaere}}, \
  and\ \bibinfo {author} {\bibfnamefont {R.~R.}\ \bibnamefont {Bouckaert}},\
  }\href {\doibase 10.1093/sysbio/syy050} {\bibfield  {journal} {\bibinfo
  {journal} {Syst. Biol.}\ ,\ \bibinfo {pages} {syy050}} (\bibinfo {year}
  {2018})}\BibitemShut {NoStop}%
\bibitem [{\citenamefont {Maturana~Russel}(2018)}]{Maturana:2018}%
  \BibitemOpen
  \bibfield  {author} {\bibinfo {author} {\bibfnamefont {P.}~\bibnamefont
  {Maturana~Russel}},\ }in\ \href@noop {} {\emph {\bibinfo {booktitle}
  {Bayesian Inference and Maximum Entropy Methods in Science and
  Engineering}}},\ \bibinfo {editor} {edited by\ \bibinfo {editor}
  {\bibfnamefont {A.}~\bibnamefont {Polpo}}, \bibinfo {editor} {\bibfnamefont
  {J.}~\bibnamefont {Stern}}, \bibinfo {editor} {\bibfnamefont
  {F.}~\bibnamefont {Louzada}}, \bibinfo {editor} {\bibfnamefont
  {R.}~\bibnamefont {Izbicki}}, \ and\ \bibinfo {editor} {\bibfnamefont
  {H.}~\bibnamefont {Takada}}}\ (\bibinfo  {publisher} {Springer International
  Publishing},\ \bibinfo {address} {Cham},\ \bibinfo {year} {2018})\ pp.\
  \bibinfo {pages} {211--219}\BibitemShut {NoStop}%
\bibitem [{\citenamefont {Green}(1995)}]{Green:1995}%
  \BibitemOpen
  \bibfield  {author} {\bibinfo {author} {\bibfnamefont {P.~J.}\ \bibnamefont
  {Green}},\ }\href {\doibase 10.1093/biomet/82.4.711} {\bibfield  {journal}
  {\bibinfo  {journal} {Biometrika}\ }\textbf {\bibinfo {volume} {82}},\
  \bibinfo {pages} {711} (\bibinfo {year} {1995})}\BibitemShut {NoStop}%
\bibitem [{\citenamefont {Umst\"atter}\ \emph {et~al.}(2005)\citenamefont
  {Umst\"atter}, \citenamefont {Christensen}, \citenamefont {Hendry},
  \citenamefont {Meyer}, \citenamefont {Simha}, \citenamefont {Veitch},
  \citenamefont {Vigeland},\ and\ \citenamefont {Woan}}]{Umstatter:2005}%
  \BibitemOpen
  \bibfield  {author} {\bibinfo {author} {\bibfnamefont {R.}~\bibnamefont
  {Umst\"atter}}, \bibinfo {author} {\bibfnamefont {N.}~\bibnamefont
  {Christensen}}, \bibinfo {author} {\bibfnamefont {M.}~\bibnamefont {Hendry}},
  \bibinfo {author} {\bibfnamefont {R.}~\bibnamefont {Meyer}}, \bibinfo
  {author} {\bibfnamefont {V.}~\bibnamefont {Simha}}, \bibinfo {author}
  {\bibfnamefont {J.}~\bibnamefont {Veitch}}, \bibinfo {author} {\bibfnamefont
  {S.}~\bibnamefont {Vigeland}}, \ and\ \bibinfo {author} {\bibfnamefont
  {G.}~\bibnamefont {Woan}},\ }\href {\doibase 10.1103/PhysRevD.72.022001}
  {\bibfield  {journal} {\bibinfo  {journal} {Phys. Rev. D}\ }\textbf {\bibinfo
  {volume} {72}},\ \bibinfo {pages} {022001} (\bibinfo {year}
  {2005})}\BibitemShut {NoStop}%
\bibitem [{\citenamefont {Cornish}\ and\ \citenamefont
  {Littenberg}(2015)}]{Cornish:2014}%
  \BibitemOpen
  \bibfield  {author} {\bibinfo {author} {\bibfnamefont {N.~J.}\ \bibnamefont
  {Cornish}}\ and\ \bibinfo {author} {\bibfnamefont {T.~B.}\ \bibnamefont
  {Littenberg}},\ }\href {\doibase 10.1088/0264-9381/32/13/135012} {\bibfield
  {journal} {\bibinfo  {journal} {Class. Quant. Grav.}\ }\textbf {\bibinfo
  {volume} {32}},\ \bibinfo {pages} {135012} (\bibinfo {year}
  {2015})}\BibitemShut {NoStop}%
\bibitem [{\citenamefont {Gelman}\ and\ \citenamefont
  {Meng}(1998)}]{Gelman:1998}%
  \BibitemOpen
  \bibfield  {author} {\bibinfo {author} {\bibfnamefont {A.}~\bibnamefont
  {Gelman}}\ and\ \bibinfo {author} {\bibfnamefont {X.}~\bibnamefont {Meng}},\
  }\href {\doibase 10.1214/ss/1028905934} {\bibfield  {journal} {\bibinfo
  {journal} {Statistical Science}\ }\textbf {\bibinfo {volume} {13}},\ \bibinfo
  {pages} {163} (\bibinfo {year} {1998})},\ \bibinfo {note} {cited By
  437}\BibitemShut {NoStop}%
\bibitem [{\citenamefont {Neal}(1993)}]{Neal:1993}%
  \BibitemOpen
  \bibfield  {author} {\bibinfo {author} {\bibfnamefont {R.~M.}\ \bibnamefont
  {Neal}},\ }\href@noop {} {\  (\bibinfo {year} {1993})}\BibitemShut {NoStop}%
\bibitem [{\citenamefont {Maturana~Russel}(2017)}]{Maturana:2017}%
  \BibitemOpen
  \bibfield  {author} {\bibinfo {author} {\bibfnamefont {P.}~\bibnamefont
  {Maturana~Russel}},\ }\emph {\bibinfo {title} {Bayesian inference in
  phylogenetics using Nested Sampling}},\ \href@noop {} {Ph.D. thesis},\
  \bibinfo  {school} {The University of Auckland} (\bibinfo {year}
  {2017})\BibitemShut {NoStop}%
\bibitem [{\citenamefont {Efron}(1979)}]{Efron:1979}%
  \BibitemOpen
  \bibfield  {author} {\bibinfo {author} {\bibfnamefont {B.}~\bibnamefont
  {Efron}},\ }\href {\doibase 10.1214/aos/1176344552} {\bibfield  {journal}
  {\bibinfo  {journal} {Ann. Statist.}\ }\textbf {\bibinfo {volume} {7}},\
  \bibinfo {pages} {1} (\bibinfo {year} {1979})}\BibitemShut {NoStop}%
\bibitem [{\citenamefont {Kunsch}(1989)}]{Kunsch:1989}%
  \BibitemOpen
  \bibfield  {author} {\bibinfo {author} {\bibfnamefont {H.~R.}\ \bibnamefont
  {Kunsch}},\ }\href {\doibase 10.1214/aos/1176347265} {\bibfield  {journal}
  {\bibinfo  {journal} {Ann. Statist.}\ }\textbf {\bibinfo {volume} {17}},\
  \bibinfo {pages} {1217} (\bibinfo {year} {1989})}\BibitemShut {NoStop}%
\bibitem [{\citenamefont {Lahiri}(2003)}]{Lahiri:2003}%
  \BibitemOpen
  \bibfield  {author} {\bibinfo {author} {\bibfnamefont {S.~N.}\ \bibnamefont
  {Lahiri}},\ }\href@noop {} {\emph {\bibinfo {title} {Resampling Methods for
  Dependent Data}}},\ Springer series in statistics\ (\bibinfo  {publisher}
  {Springer New York},\ \bibinfo {address} {New York, NY},\ \bibinfo {year}
  {2003})\BibitemShut {NoStop}%
\bibitem [{\citenamefont {Hannam}\ \emph {et~al.}(2014)\citenamefont {Hannam}
  \emph {et~al.}}]{Hannam:2013oca}%
  \BibitemOpen
  \bibfield  {author} {\bibinfo {author} {\bibfnamefont {M.}~\bibnamefont
  {Hannam}} \emph {et~al.},\ }\href {\doibase 10.1103/PhysRevLett.113.151101}
  {\bibfield  {journal} {\bibinfo  {journal} {\prl}\ }\textbf {\bibinfo
  {volume} {113}},\ \bibinfo {pages} {151101} (\bibinfo {year} {2014})},\
  \Eprint {http://arxiv.org/abs/1308.3271} {arXiv:1308.3271 [gr-qc]}
  \BibitemShut {NoStop}%
\bibitem [{\citenamefont {Fan}\ \emph {et~al.}(2011)\citenamefont {Fan},
  \citenamefont {Wu}, \citenamefont {Chen}, \citenamefont {Kuo},\ and\
  \citenamefont {Lewis}}]{fan:2011}%
  \BibitemOpen
  \bibfield  {author} {\bibinfo {author} {\bibfnamefont {Y.}~\bibnamefont
  {Fan}}, \bibinfo {author} {\bibfnamefont {R.}~\bibnamefont {Wu}}, \bibinfo
  {author} {\bibfnamefont {M.-H.}\ \bibnamefont {Chen}}, \bibinfo {author}
  {\bibfnamefont {L.}~\bibnamefont {Kuo}}, \ and\ \bibinfo {author}
  {\bibfnamefont {P.~O.}\ \bibnamefont {Lewis}},\ }\href {\doibase
  10.1093/molbev/msq224} {\bibfield  {journal} {\bibinfo  {journal} {Mol. Biol.
  Evol.}\ }\textbf {\bibinfo {volume} {28}},\ \bibinfo {pages} {523} (\bibinfo
  {year} {2011})}\BibitemShut {NoStop}%
\end{thebibliography}%

\end{document}